\documentclass{aa}

\usepackage{graphicx}
\usepackage{txfonts}
\usepackage{amsmath}
\usepackage{booktabs}
\usepackage{multirow}
\usepackage{array}
\usepackage{threeparttable}
\usepackage{makecell}
\usepackage{adjustbox}
\usepackage{subfigure}
\usepackage[colorlinks=true,allcolors=blue]{hyperref}
\usepackage{orcidlink}

\newcommand{\civ}{C\,\textsc{iv}\,}

\newcommand{\mgii}{Mg\,\textsc{ii}\,}
\newcommand{\pa}{P$\alpha$\,}
\newcommand{\pb}{P$\beta$\,}
\newcommand{\ha}{H$\alpha$\,}
\newcommand{\hb}{H$\beta$\,}

\begin{document}

\title{Broad Paschen lines as black hole mass tracers}
\subtitle{A reverberation mapping calibration}
\titlerunning{Broad Paschen lines as black hole mass tracers}

\author{
S. Sun\orcidlink{0000-0002-1234-552X}\inst{1,2,3}
      \corrauth{sxsun@stu.pku.edu.cn}
\and S. E. I. Bosman\orcidlink{0000-0001-8582-7012}\inst{1,4}
      \email{bosman@thphys.uni-heidelberg.de}
\and F. B. Davies\orcidlink{0000-0003-0821-3644}\inst{4}\email{davies@mpia.de}
\and K. Protu\v{s}ov\'a\orcidlink{0009-0008-2205-7725}\inst{1}
      \email{protusova@thphys.uni-heidelberg.de}
\and B. Spina\orcidlink{0000-0003-1634-1283}\inst{1}\email{spina@thphys.uni-heidelberg.de}
\and L. Jiang\orcidlink{0000-0003-4176-6486}\inst{2,3}\email{jiangKIAA@pku.edu.cn}
\and D. Jiang\orcidlink{0009-0003-6747-2221}\inst{2,3}\email{jiangdy@stu.pku.edu.cn}
}
\authorrunning{Sun et al.}

\institute{Institute for Theoretical Physics, Heidelberg University, Philosophenweg 12, 69120 Heidelberg, Germany
\and Department of Astronomy, School of Physics, Peking University, Beijing 100871, China
\and Kavli Institute for Astronomy and Astrophysics, Peking University, Beijing 100871, China
\and Max-Planck-Institut f\"ur Astronomie, K\"onigstuhl 17, 69117 Heidelberg, Germany}

\date{Received XXX / Accepted XXX}

\abstract
{Single-epoch (SE) black hole mass estimators commonly rely on rest-frame ultraviolet or optical broad lines. Their use can be limited by dust attenuation or wavelength coverage. Hydrogen Paschen lines provide longer-wavelength tracers, but previous calibrations often included masses inferred from other SE relations.}
{We calibrate black hole mass estimators based on broad Paschen P$\alpha$ and P$\beta$ emission against black hole masses from reverberation mapping (RM). We also assess how the line profile and luminosity measurements affect the calibrated relations.}
{We obtained near-infrared spectra of 27 quasars from the Sloan Digital Sky Survey Reverberation Mapping sample with the Large Binocular Telescope (LBT) Utility Cameras in the Infrared (LUCI). We fitted the continuum-subtracted Paschen profiles with multi-component and single-Gaussian models and measured line luminosities and full widths at half maximum (FWHMs). We used rest-frame 5100~\AA\ luminosities from archival optical spectra as alternative continuum measurements. We combined the LBT sample with archival Paschen measurements of objects with RM-based masses and fitted $\log(M_{\rm BH}/M_{\odot})=a+b\log(L/10^{41}\,{\rm erg\,s^{-1}})+c\log({\rm FWHM}/10^3\,{\rm km\,s^{-1}})$, including intrinsic scatter. Here, $a$ is the normalization, and $b$ and $c$ are the luminosity and FWHM slopes.}
{The fiducial relations use photometrically corrected Paschen line luminosities and the FWHM of the combined broad profile from the multi-component fit. For P$\alpha$, we obtain $a=6.86^{+0.24}_{-0.25}$, $b=0.47^{+0.11}_{-0.10}$, and $c=1.24^{+0.49}_{-0.48}$, with an intrinsic scatter of $0.28^{+0.08}_{-0.06}$ dex. For P$\beta$, we obtain $a=7.00\pm0.24$, $b=0.48^{+0.10}_{-0.09}$, and $c=0.91^{+0.46}_{-0.47}$, with an intrinsic scatter of $0.32^{+0.08}_{-0.06}$ dex. The fitted scatters for the multi-component, single-Gaussian, and constrained slope relations overlap within their uncertainties.}
{The P$\alpha$ and P$\beta$ relations anchored to RM masses provide complementary mass estimators when optical or ultraviolet broad line measurements are affected by attenuation or unavailable. They should be applied within the parameter ranges of the calibration sample and remain subject to the systematic uncertainty associated with the RM virial factor.}

%\keywords{galaxies: active -- quasars: emission lines -- galaxies: nuclei -- black hole physics -- infrared: galaxies}

\maketitle
\nolinenumbers

\section{Introduction} \label{sec:intro}

Supermassive black holes reside at the centers of most massive galaxies. When actively accreting, they can power quasars. Empirical correlations between the mass of the central black hole ($M_{\rm BH}$) and host galaxy properties \citep{Magorrian1998, Ferrarese2000, Gebhardt2000, Kormendy2013} are commonly discussed in the context of black hole and galaxy coevolution and active galactic nucleus (AGN) feedback \citep{Fabian2012, Heckman2014}. Measurements of $M_{\rm BH}$ across cosmic time therefore help characterize black hole growth and its relation to galaxy evolution.

Black hole masses can be estimated through several observational approaches. In nearby galaxies, spatially resolved stellar and gas dynamics, including measurements of circumnuclear water megamaser disks, constrain the mass of the central compact object \citep{Kormendy2013,Kuo2011}. Velocity-resolved spectroastrometry constrains broad line region (BLR) sizes and kinematics from wavelength-dependent photocenter shifts. Such measurements can use either single-aperture spectroscopy assisted by adaptive optics \citep{Stern2015,Bosco2021} or long-baseline interferometry in the near-infrared (NIR).
Differential measurements with the GRAVITY instrument on the Very Large Telescope Interferometer have resolved BLR rotation and provided dynamical black hole mass constraints in AGNs from the local Universe to $z\simeq2.3$ \citep{GRAVITY2018,GRAVITY2024,Abuter2024}. Strong gravitational lensing combined with James Webb Space Telescope (JWST) integral-field spectroscopy has also enabled a dynamical measurement of a central point mass for a little red dot (LRD) at $z=7.04$ \citep{Juodzbalis2026}.
Gravitational waves provide mass information in a different observational regime. For coalescing compact binaries, gravitational-wave waveforms encode the component and remnant masses \citep{Abbott2016}. At nanohertz frequencies, pulsar timing measurements of a stochastic background constrain possible populations of supermassive black hole binaries statistically rather than measuring individual quasar black hole masses \citep{Agazie2023}.

For AGNs with broad emission lines whose central regions are not spatially resolved, reverberation mapping (RM) measures the BLR scale from the delay between continuum variability and the response of broad emission lines \citep{Blandford1982, Peterson1993, Peterson2004}. Assuming virial equilibrium for the BLR, $M_{\rm BH}$ can be estimated from the BLR size ($R_{\rm BLR}$) and a characteristic velocity ($\Delta V$) inferred from the broad line width:
\begin{equation}
    M_{\rm BH} = f \frac{R_{\mathrm{BLR}} \, \Delta V^2}{G} ,
\end{equation}
where $G$ is the gravitational constant and the virial factor $f$ accounts for the geometry, inclination, and kinematics of the BLR as well as the adopted definition of line width. RM determines the virial product, $R_{\mathrm{BLR}}\Delta V^2/G$, whereas conversion to an absolute black hole mass requires either an individual BLR dynamical model or an externally calibrated value of $f$.

The RM mass is therefore a virial estimate dependent of modeling. For the Sloan Digital Sky Survey (SDSS) Reverberation Mapping (SDSS-RM) catalog, \citet{Shen_rm_2024ApJS..272...26S} calibrated a mean virial factor using 30 low-redshift RM AGNs with masses from BLR dynamical models. They obtained $\langle\log f\rangle=0.62\pm0.07$ for the line dispersion measured from the root-mean-square (rms) spectrum, with an intrinsic scatter of $0.31\pm0.07$ dex among objects. This scatter adds an uncertainty of approximately a factor of two to an individual RM mass and propagates into calibrations anchored to RM masses.

Spectroscopic RM campaigns have extended lag measurements across a wider range of AGN properties. The Super-Eddington Accreting Massive Black Hole campaign targeted AGNs with high accretion rates and reported H$\beta$ lags shorter than those predicted by the canonical relation between radius and luminosity for some targets \citep{Du2014SEAMBH, Wang2014SEAMBH, Du2018SEAMBH}. Multi-object programs include SDSS-RM, which has monitored a large quasar sample over a multi-year baseline \citep{Shen2015,Grier2017,Homayouni2020,Shen_rm_2024ApJS..272...26S}, and the RM program of the Australian Dark Energy Survey, for which \citet{Yu2023OzDES} reported 25 Mg\,{\sc ii} lags from six years of monitoring. Subject to uncertainty in the virial factor, these RM samples provide empirical mass references for calibrating secondary estimators.

Because RM requires repeated monitoring, single-epoch (SE) virial estimators are commonly used for large samples and high-redshift quasars \citep[e.g.,][]{Vestergaard2006, McLure2004}. These estimators use the empirical relation between radius and luminosity ($R-L$) \citep{Kaspi2000, Bentz2013} to infer $R_{\rm BLR}$ from an AGN continuum or emission line luminosity. SE estimators commonly use rest-frame ultraviolet and optical broad lines, including H$\beta$, Mg\,{\sc ii}, and C\,{\sc iv} \citep{Vestergaard2006, Pan_2025}. For dust-reddened quasars, or when these lines are shifted beyond optical wavelength coverage, longer-wavelength tracers provide complementary measurements.
AGNs that are reddened by dust or obscured may represent a stage in quasar evolution between ultraluminous infrared galaxies (ULIRGs) and unobscured quasars \citep{Sanders1988, Urrutia2009, Glikman2013}. These systems contribute to cosmic black hole growth \citep{Soltan1982, Fabian1999, Inayoshi2020, FanXiaohui2023}, while dust attenuation introduces additional uncertainty into measurements based on optical tracers \citep{Bosman2024, Bosman_2025, dust_att_lrd_2025NatAs...9.1732S}.

The NIR Paschen series, particularly \pa ($\lambda 1.875\ \mu{\rm m}$) and \pb ($\lambda 1.282\ \mu{\rm m}$), is less affected by dust attenuation than optical hydrogen lines. Several studies have calibrated $M_{\rm BH}$ estimators based on these lines \citep{Kim_2010, LaFranca2015, Kim_2015, Ricci2017}, and NIR spectroscopy has been used to identify and characterize broad line emission in obscured AGNs \citep{Landt2008, Landt2011, Onori2017, denBrok2022}. Estimators based on Paschen lines may also be applicable to LRDs, some of which show broad emission lines that have been interpreted as evidence for accreting black holes \citep[e.g.,][]{Kocevski2023, Maiolino2024, Greene2024, Matthee2024}. The lower attenuation of Paschen lines and their lack of resonant scattering motivate their use as additional broad line tracers.

Previous Paschen line calibrations used samples that included black hole masses inferred from Balmer line SE estimators \citep{Kim_2010, Kim_2015}. Calibration against RM masses avoids this intermediate Balmer SE estimate, but remains subject to uncertainty in the RM virial factor and should not be interpreted as a calibration against exact black hole masses. Published analyses have also used different profile models, including single- and multi-component Gaussian fits, which can yield different line width measurements. The archival sample used in this work contains 11 quasars with both Paschen line measurements and RM-based black hole masses.

In this work, we calibrate SE black hole mass estimators based on \pa\ and \pb\ against RM masses. We use NIR spectra of 27 SDSS-RM quasars obtained with the Large Binocular Telescope (LBT) Utility Cameras in the Infrared (LUCI) and examine how single- and multi-component profile models affect the fitted relations. We provide the resulting calibrations for estimating black hole masses when Paschen lines are available, including in reddened and obscured quasars.

\begin{table*}[t]
\caption{Paschen line luminosity and FWHM measurements from multi-component fits for the SDSS-RM sample.}
\label{tab:paschen_params}
\centering
\begin{tabular}{lcccccc}
\hline\hline
RMID & $z$ & $\log L_{\rm P\beta}$ & $\log L_{\rm P\alpha}$ &
FWHM$_{\rm P\beta}$ & FWHM$_{\rm P\alpha}$ & $\log M_{\rm BH,RM}$ \\
 & & (erg s$^{-1}$) & (erg s$^{-1}$) &
(km s$^{-1}$) & (km s$^{-1}$) & ($M_\odot$) \\
\hline
118-SDSS-RM & 0.72 & $42.50^{+0.08}_{-0.14}$ & $\ldots$ & $3188 \pm 204$ & $\ldots$ & $8.27 \pm 0.02$ \\
126-SDSS-RM & 0.19 & $41.00^{+0.05}_{-0.05}$ & $41.24^{+0.05}_{-0.04}$ & $1659 \pm 114$ & $3375 \pm 140$ & $7.56 \pm 0.05$ \\
184-SDSS-RM & 0.19 & $41.51^{+0.06}_{-0.05}$ & $41.56^{+0.07}_{-0.07}$ & $2122 \pm 127$ & $3024 \pm 729$ & $8.15 \pm 0.17$ \\
272-SDSS-RM & 0.26 & $41.90^{+0.05}_{-0.05}$ & $41.96^{+0.05}_{-0.05}$ & $3162 \pm 123$ & $3201 \pm 152$ & $7.52 \pm 0.08$ \\
316-SDSS-RM & 0.68 & $42.88^{+0.12}_{-0.11}$ & $\ldots$ & $2700 \pm 143$ & $\ldots$ & $8.17 \pm 0.12$ \\
329-SDSS-RM & 0.72 & $42.87^{+0.10}_{-0.09}$ & $\ldots$ & $3014 \pm 104$ & $\ldots$ & $8.13 \pm 0.03$ \\
766-SDSS-RM & 0.16 & $\ldots$ & $41.36^{+0.05}_{-0.05}$ & $\ldots$ & $1710 \pm 95$ & $6.94 \pm 0.13$ \\
769-SDSS-RM & 0.19 & $\ldots$ & $40.84^{+0.05}_{-0.04}$ & $\ldots$ & $3052 \pm 233$ & $7.22 \pm 0.20$ \\
775-SDSS-RM & 0.17 & $\ldots$ & $41.65^{+0.03}_{-0.04}$ & $\ldots$ & $2361 \pm 64$ & $\ldots$ \\
776-SDSS-RM & 0.12 & $\ldots$ & $40.81^{+0.05}_{-0.05}$ & $\ldots$ & $2330 \pm 132$ & $7.67 \pm 0.08$ \\
781-SDSS-RM & 0.26 & $41.25^{+0.13}_{-0.08}$ & $41.44^{+0.05}_{-0.17}$ & $2704 \pm 444$ & $2651 \pm 244$ & $7.21 \pm 0.08$ \\
822-SDSS-RM & 0.29 & $41.10^{+0.10}_{-0.11}$ & $\ldots$ & $1925 \pm 288$ & $\ldots$ & $6.65 \pm 0.08$ \\
\hline
\end{tabular}
\tablefoot{RMID denotes the SDSS-RM identifier. The listed luminosities are $\log L$ in units of
erg s$^{-1}$, and the FWHM values are in units of km s$^{-1}$ and
are corrected for instrumental resolution. Line luminosities and FWHM
values are measured from the broad Paschen profile using the
multi-component Gaussian model. Ellipses indicate that the object was
not observed in that band, the line was not detectable, or no RM black
hole mass is available. The line luminosities include the NIR photometric correction for slit losses; their uncertainties combine the profile fitting and photometric scaling terms.}
\end{table*}

The remainder of this paper is organized as follows. Sect.~\ref{sec:sample} describes the SDSS-RM target selection, LBT/LUCI observations, data reduction, and measurements of the Paschen line and optical continuum properties. Sect.~\ref{sec:result} presents the regression method and the resulting mass calibrations based on \pa\ and \pb. Sect.~\ref{sec:discussion} discusses the fiducial estimators, compares them with previous Paschen relations, and considers sources without detected broad Paschen components. Sect.~\ref{sec:summary} summarizes the results. Throughout this paper, we use AB magnitudes and adopt a flat $\Lambda$ cold dark matter ($\Lambda$CDM) cosmology with $H_0 = 70\, \mathrm{km}\,\mathrm{s}^{-1}\,\mathrm{Mpc}^{-1}$, $\Omega_{\Lambda}=0.7$, and $\Omega_m=0.3$.

\section{Sample and reduction}  \label{sec:sample}

\subsection{Sample and LBT observations}

We selected our targets from SDSS-RM, which monitored 849 quasars over a 10-year baseline and provides lag measurements, RM-based black hole masses ($M_{\rm RM}$), and optical spectral parameters \citep{Shen_rm_2024ApJS..272...26S}.
Among the 849 quasars, 90 have \pa\ and/or \pb\ redshifted into the wavelength coverage of the LUCI NIR spectrographs \citep{Seifert2003} on LBT. We selected the 30 brightest quasars from this subset for NIR spectroscopy (program MPIA-2024A-004).
Most of the observations were obtained under suboptimal conditions, with relatively poor seeing and, in some cases, thin cloud.

We observed the targets for a total of 11~hr using both LUCI1 and LUCI2. We used an A--B--B--A (ABBA) dither sequence with an individual exposure time of 225~s, incurring 378~s of total overhead per target.
We used the high-resolution H- and K-band gratings to resolve the Paschen profiles and selected the central wavelengths to cover both \pa\ and \pb\ for 16 quasars.

\begin{figure*}[t]
\centering
\includegraphics[width=0.85\textwidth]{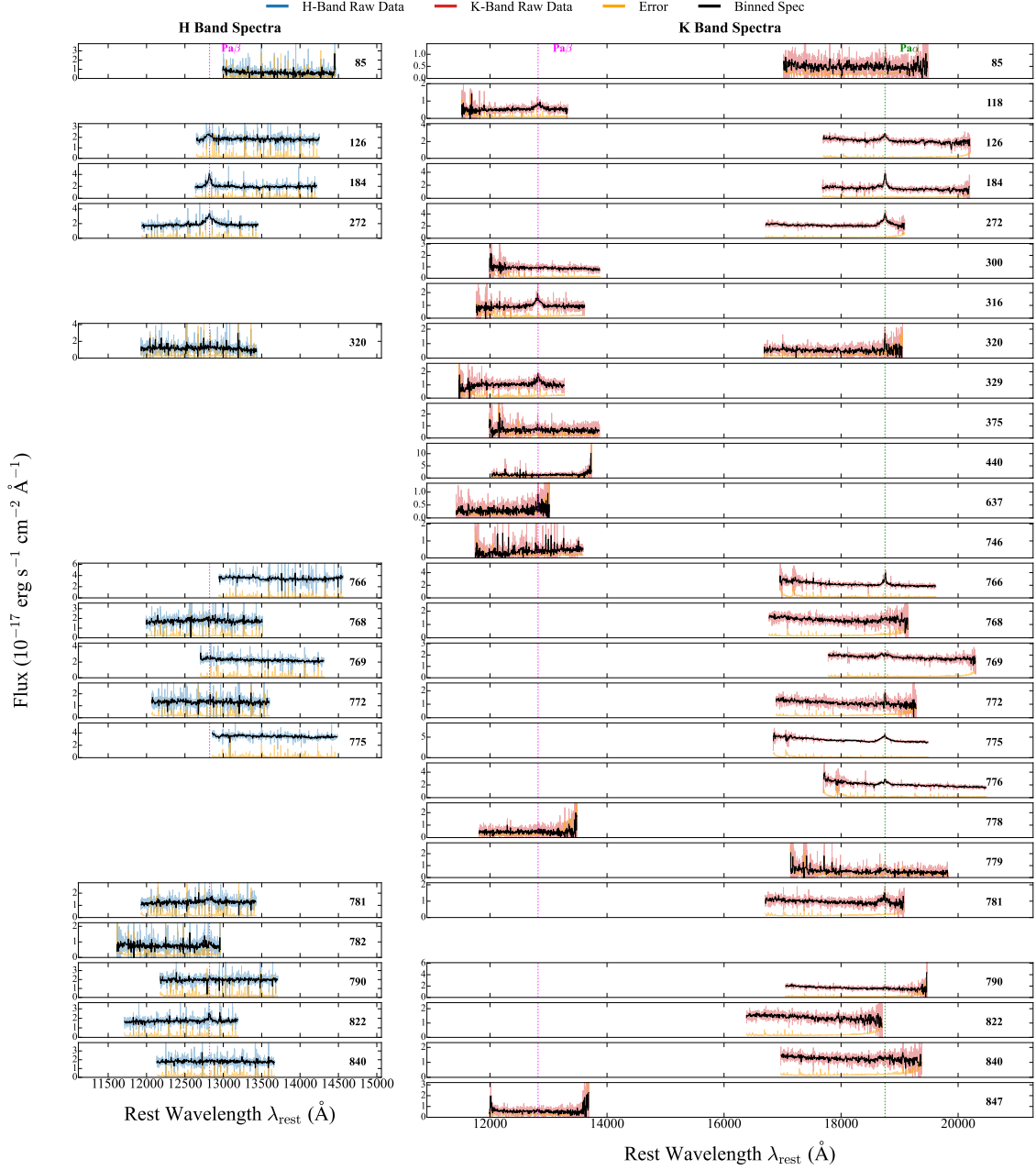}
\caption{NIR spectral atlas for the 27 SDSS-RM quasars with reduced LBT/LUCI spectra.
        Blue and red lines show the coadded H- and K-band spectra, respectively, and black lines show the binned spectra for clarity.
        Magenta and green dashed lines mark the \pa\ and \pb\ wavelengths.
        Wavelengths are shown in the rest frame using the SDSS-RM redshifts.
        %The y axis shows flux density $f_{\lambda}$ in units of $10^{-17}\ {\rm erg\,s^{-1}\,cm^{-2}}\,\text{\AA}^{-1}$.
\label{spec_atlas}}
\end{figure*}

We obtained data for 30 quasars, although grating mistunings or poor observing conditions affected some spectra. The mistunings removed H-band coverage for four objects and K-band coverage for RMID 822, whose H-band spectrum contains a detected broad \pb line. The signal-to-noise ratio (S/N) was insufficient for the intended measurements in both bands for two targets and primarily in the K band for four targets. RMID 769 shows a possible \pa absorption feature rather than a broad emission component.
We detected both \pa\ and \pb\ in four quasars. After excluding spectra without a measurable broad Paschen component, 26 quasars remain for line fitting. Only the subset with an RM mass and a line measurement satisfying the adopted quality criteria enters each calibration.
Figure~\ref{spec_atlas} shows the coadded spectra, and Table~\ref{tab:paschen_params} lists the measured Paschen line properties.

\subsection{Data reduction and spectral measurements} \label{sec:sample_reduction}

We reduced the LBT/LUCI NIR spectra with \texttt{PypeIt} \citep{Pypeit_2020JOSS....5.2308P}.
For each H- or K-band setup, the reduction included processing of calibration frames, slit tracing, wavelength calibration, sky subtraction using the ABBA dither pattern, one-dimensional extraction, flux calibration, and coaddition.
We corrected the coadded spectra for telluric absorption and foreground Galactic extinction and shifted them to the rest frame using the SDSS-RM redshifts.
We masked regions with strong atmospheric residuals or anomalous flux peaks during the subsequent spectral fitting.

Some SDSS-RM quasars show broad \ha or \hb emission in their optical spectra but no measurable broad \pa or \pb component in the reduced LBT spectra. We excluded spectra without a measurable broad Paschen component from the corresponding calibration sample. We discuss possible causes of the difference between the optical and NIR line detections in Sect.~\ref{sec:paschen_nondetections}.

\begin{figure*}[t]
\centering
\includegraphics[width=0.83\textwidth]{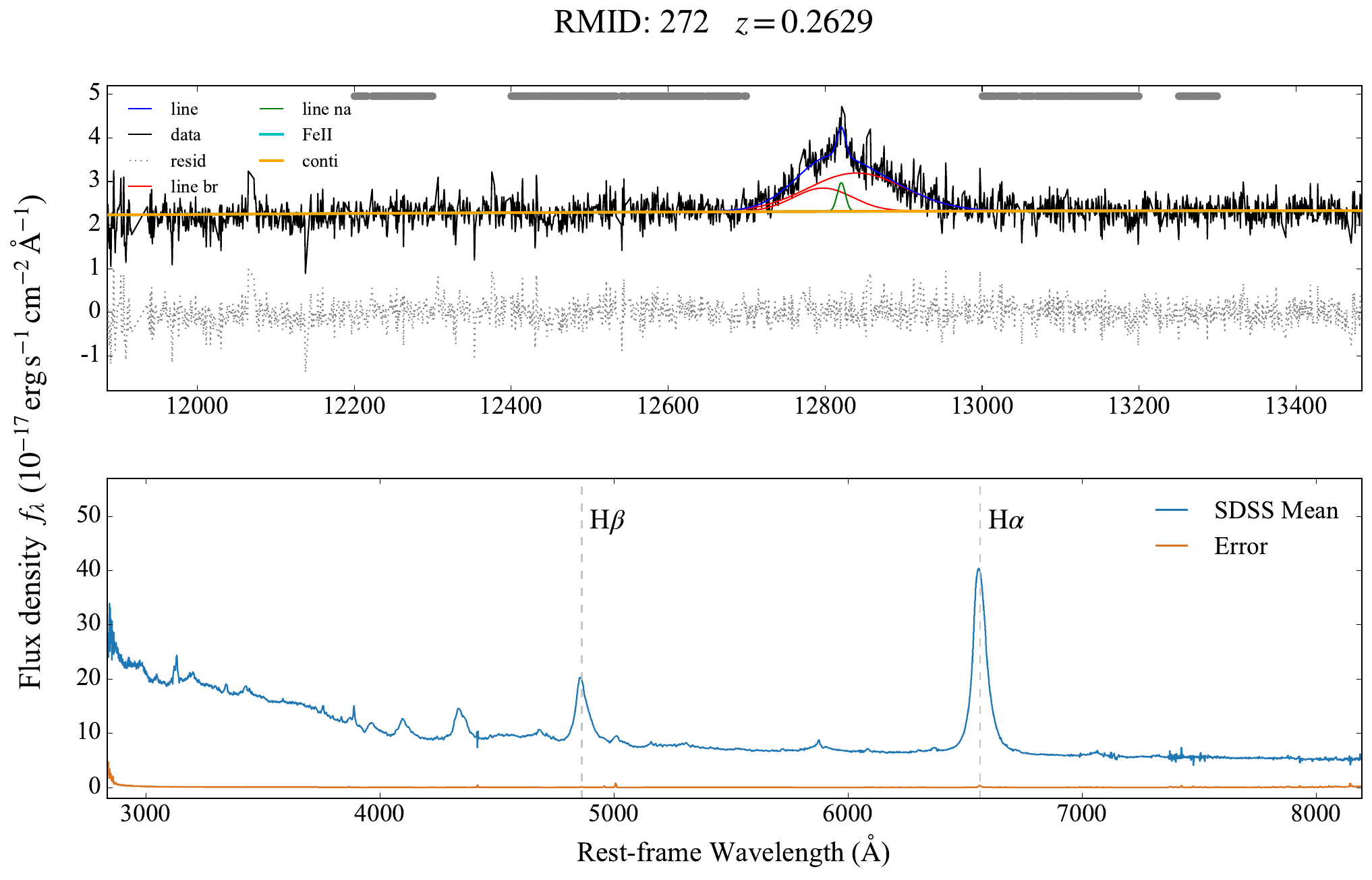}
\caption{Example spectral fit for SDSS-RM-272.
The upper panel shows the LBT H-band spectrum around the rest-frame \pb\ line (black), the best-fitting \texttt{PyQSOFit} model (blue), the broad and narrow Gaussian components (red and green), and the residuals (gray dotted line).
The lower panel shows the archival SDSS-RM mean optical spectrum.
\label{fig:example_fitting_res}}
\end{figure*}

We measured the Paschen line properties with \texttt{PyQSOFit} \citep{pyqsofit_2018ascl.soft09008G, pyqsofit_2019ApJS..241...34S}. 
For each reduced NIR spectrum, we fitted the local continuum with a power law and a third-order polynomial, using continuum windows chosen to avoid strong emission lines and telluric residuals.
After subtracting the best-fit continuum, we modeled the \pa\ and \pb\ emission line profiles with Gaussian components. 
Figure~\ref{fig:example_fitting_res} shows an example fit.

We adopted two line width measurements for each Paschen line. 
We derived our full width at half maximum (FWHM) fiducial measurement, $\mathrm{FWHM}_{\rm multi}$, from a multi-component model with two broad Gaussian components and, when required, one narrow Gaussian component to achieve a reasonable and reliable fit. The $\mathrm{FWHM}_{\rm multi}$ only accounts for the broad components.
This prescription is similar to the multi-Gaussian decomposition used in previous Paschen line calibrations \citep[e.g.,][]{Kim_2010, Kim_2015}. 
For comparison, we also fitted each Paschen profile with a single Gaussian and report the corresponding width as $\mathrm{FWHM}_{\sigma}$. 
This second measurement provides a simpler description that may be useful for lower-S/N or lower-resolution spectra, where a detailed decomposition of the broad line profile is not always justified.
We measured the line luminosities, $L_{\rm P\alpha}$ and $L_{\rm P \beta}$, by integrating the fitted broad Paschen components after continuum subtraction.
We corrected all reported FWHM values for instrumental broadening.
We estimated the line width and luminosity uncertainties by propagating the Markov chain Monte Carlo (MCMC) samples from \texttt{PyQSOFit}. The reported uncertainty is half the difference between the 16th and 84th percentiles for each property.
During the fitting, we masked strong sky line residuals and pixels with flux uncertainties larger than three times the median uncertainty within the fitting window. 
We visually inspected all fits and used only spectra that met the adopted broad line measurement criteria in the calibration.

To correct for slit losses through the 0.5~arcsec LUCI slit, we rescaled each H- and K-band spectrum to external NIR photometry. We treated the slit loss as a multiplicative factor within each spectral setup. We used Two Micron All Sky Survey (2MASS) $JHK_s$ photometry where available \citep{Skrutskie2006}. For the three objects without a 2MASS counterpart, we used forced spectrophotometry from Quick Release 2 of the Spectro-Photometer for the History of the Universe, Epoch of Reionization and Ices Explorer (SPHEREx), obtained with the NASA/IPAC Infrared Science Archive Spectrophotometry Tool \citep{Bock2026,Akeson2025}.
For each setup, we compared the external photometric flux density with the LBT continuum after masking emission lines and applied the resulting scale factor to the integrated Paschen line flux. The correction factors range from 1.19 to 3.67 (0.08--0.56 dex). We propagated their wavelength-dependent scatter in quadrature with the line fitting uncertainty. Source variability may also affect the scale factors because the photometry and spectroscopy are not simultaneous but this effect is marginal.
We compared the SPHEREx and 2MASS correction factors for the 12 line measurements with both estimates. Their median difference is 0.01 dex. After excluding RMID~776, for which the SPHEREx fit quality statistic indicates a poor forced fit, the median difference is much smaller than 0.01 dex and the rms difference is 0.10 dex. We therefore use 2MASS as the primary photometric scale where available and SPHEREx for the four line measurements without 2MASS coverage.

We also measured the rest-frame 5100~\AA\ optical continuum luminosity from the archival SDSS-RM mean spectra \citep{Shen_rm_2024ApJS..272...26S}. 
We fitted these spectra with \texttt{PyQSOFit} using a similar continuum model.
When present, the main broad optical emission lines, including \mgii, \hb, and \ha, were modeled with multiple Gaussian components.
We used the resulting AGN continuum luminosity, $L_{5100}$, as an alternative luminosity term in the SE mass calibrations.
%Because these optical spectra have an independent flux calibration, the LUCI correction for slit losses was not applied to $L_{5100}$.

\begin{figure*}[t]
\centering
\includegraphics[width=0.79\textwidth]{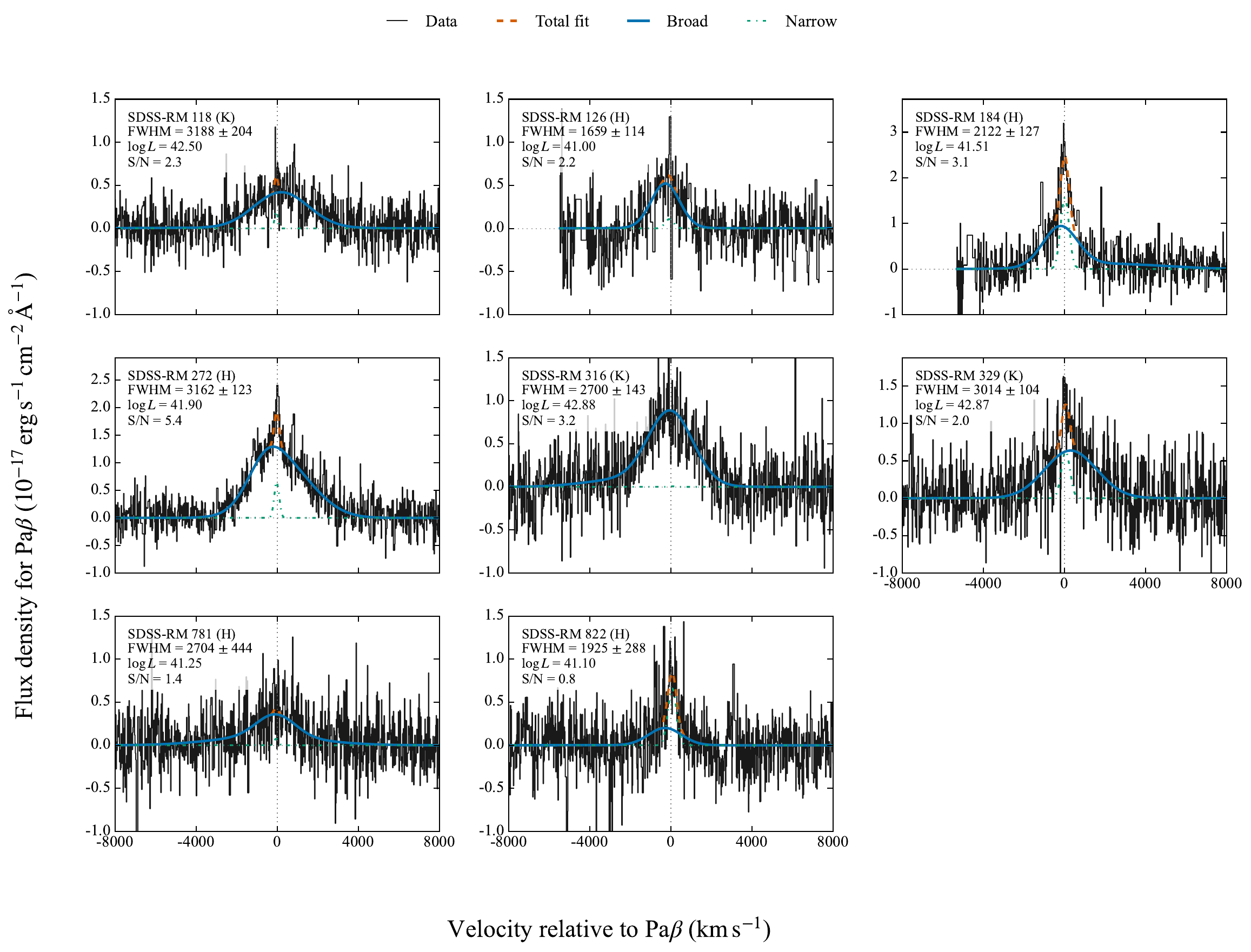}
\caption{Atlas in velocity space of the \pb line profile fits used for the black hole mass calibration.
    Only SDSS-RM sources that passed the fit quality and visual inspection criteria are shown.
    Each panel presents the LBT spectrum after continuum subtraction in black, the total best-fit line model as an orange dashed curve, the combined broad component in blue, and the narrow component as a green dash-dotted curve when included in the fit.
    Velocities are measured relative to the vacuum \pb rest wavelength, and the same colors and line styles are used in all panels.}
\label{fig:pa_beta_calibration_atlas}
\end{figure*}

\begin{figure*}[t]
\centering
\includegraphics[width=0.79\textwidth]{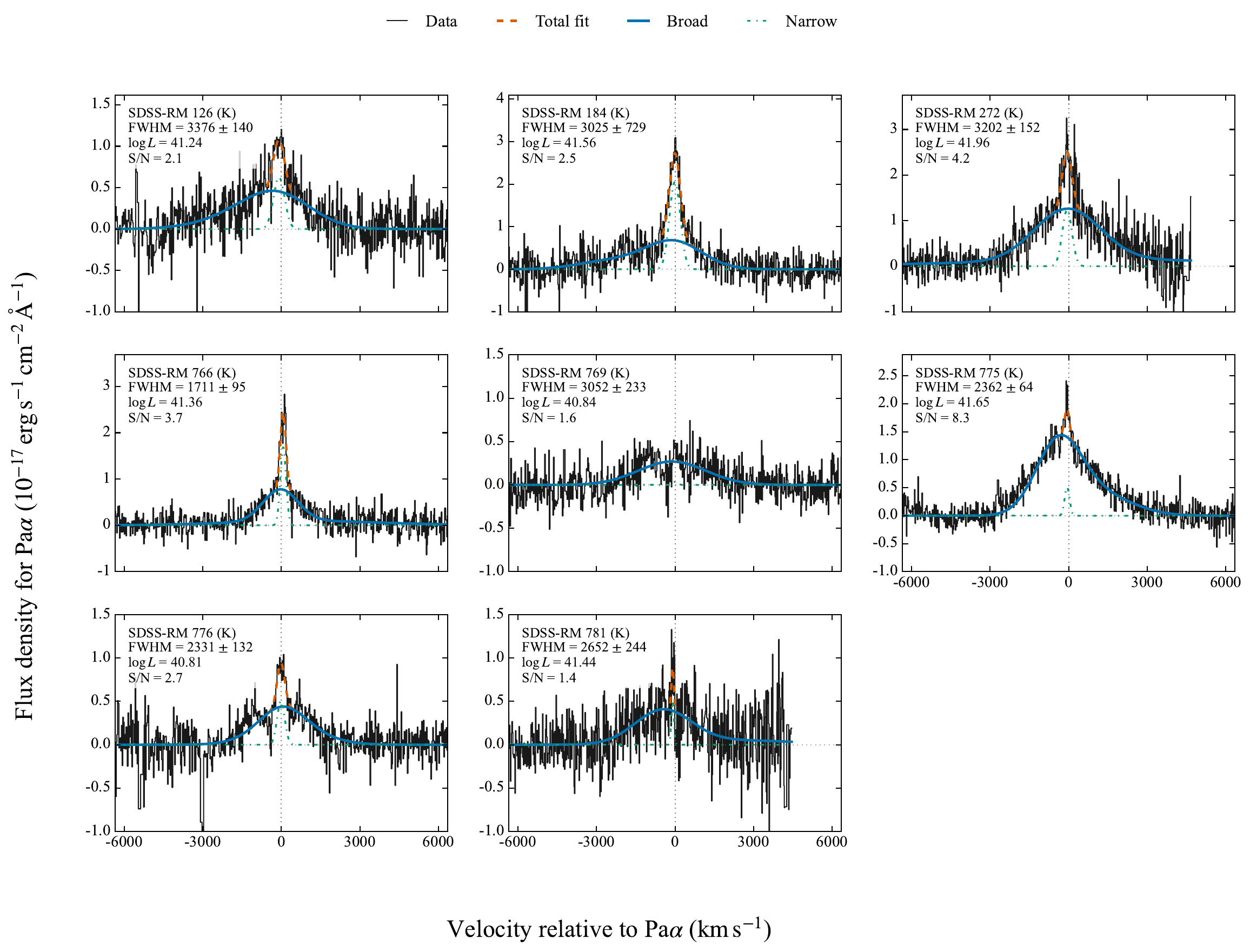}
\caption{Same as Fig.~\ref{fig:pa_beta_calibration_atlas}, but for \pa.
    Only sources that pass the adopted calibration selection are shown, and velocities are measured relative to the vacuum \pa\ rest wavelength.}
\label{fig:pa_alpha_calibration_atlas}
\end{figure*}

Figures~\ref{fig:pa_beta_calibration_atlas} and~\ref{fig:pa_alpha_calibration_atlas} show the continuum-subtracted Paschen profiles and best-fitting components used in the calibration.

We summarize the Paschen line measurements in Table~\ref{tab:paschen_params}.
We adopted the RM black hole masses from \citet{Shen_rm_2024ApJS..272...26S}.
When multiple RM masses were available for the same object, we prioritized them in the order \hb, \ha, \mgii, and \civ.
To extend the dynamic range of the calibration and maintain consistency with previous Paschen line work, we also included the \citet{Kim_2010} objects that have both Paschen line measurements and black hole masses from RM.

\section{Calibration results} \label{sec:result}

We calibrated SE black hole mass estimators of the form
\begin{equation}
\label{eq:paschen_mbh_calibration}
\begin{split}
    \log \left( \frac{M_{\rm BH}}{M_{\odot}} \right)
    =\ & a
    + b \log \left( \frac{L}{10^{41}\ {\rm erg\ s^{-1}}} \right) \\
    &+ c \log \left( \frac{\rm FWHM}{10^{3}\ {\rm km\ s^{-1}}} \right),
\end{split}
\end{equation}
where $L$ is either the Paschen line luminosity or the rest-frame 5100~\AA\ continuum luminosity, and FWHM is measured from either the multi-component broad line model or the single-Gaussian model described in Sect.~\ref{sec:sample_reduction}. 
We fit the coefficients $a$, $b$, and $c$, together with the intrinsic scatter $\sigma_{\rm int}$, using \texttt{emcee} \citep{emcee_2013PASP..125..306F}.

For each object, the likelihood includes the uncertainty in the RM black hole mass and the propagated uncertainties in luminosity and FWHM. 
For a data point $i$, we define the combined uncertainty as
\begin{equation}
    s_i^2 =
    \delta y_i^2
    + b^2 \delta x_{L,i}^2
    + c^2 \delta x_{{\rm FWHM},i}^2
    + \sigma_{\rm int}^2 ,
\end{equation}
where $y_i=\log(M_{\rm BH,RM}/M_{\odot})$, $x_{L,i}=\log(L/10^{41}\ {\rm erg\ s^{-1}})$, and $x_{{\rm FWHM},i}=\log({\rm FWHM}/10^{3}\ {\rm km\ s^{-1}})$. The terms $\delta y_i$, $\delta x_{L,i}$, and $\delta x_{{\rm FWHM},i}$ are the corresponding $1\sigma$ uncertainties in logarithmic space. 
For asymmetric uncertainties, we use the mean of the upper and lower errors after converting them to logarithmic units. 
The term $\sigma_{\rm int}$ represents the intrinsic scatter of the calibration relation, also in dex.
The log-likelihood is then
\begin{equation}
    \ln \mathcal{L}
    =
    -\frac{1}{2}
    \sum_i
    \left[
    \frac{(y_i-y_{{\rm model},i})^2}{s_i^2}
    + \ln(s_i^2)
    \right],
\end{equation}
where 
$y_{{\rm model},i}=a+b x_{L,i}+c x_{{\rm FWHM},i}$.
We used uniform priors of $-20<a<20$, $-10<b,c<10$, and $0<\sigma_{\rm int}<3$.

\begin{figure}[t]
\centering
\includegraphics[width=0.95\columnwidth]{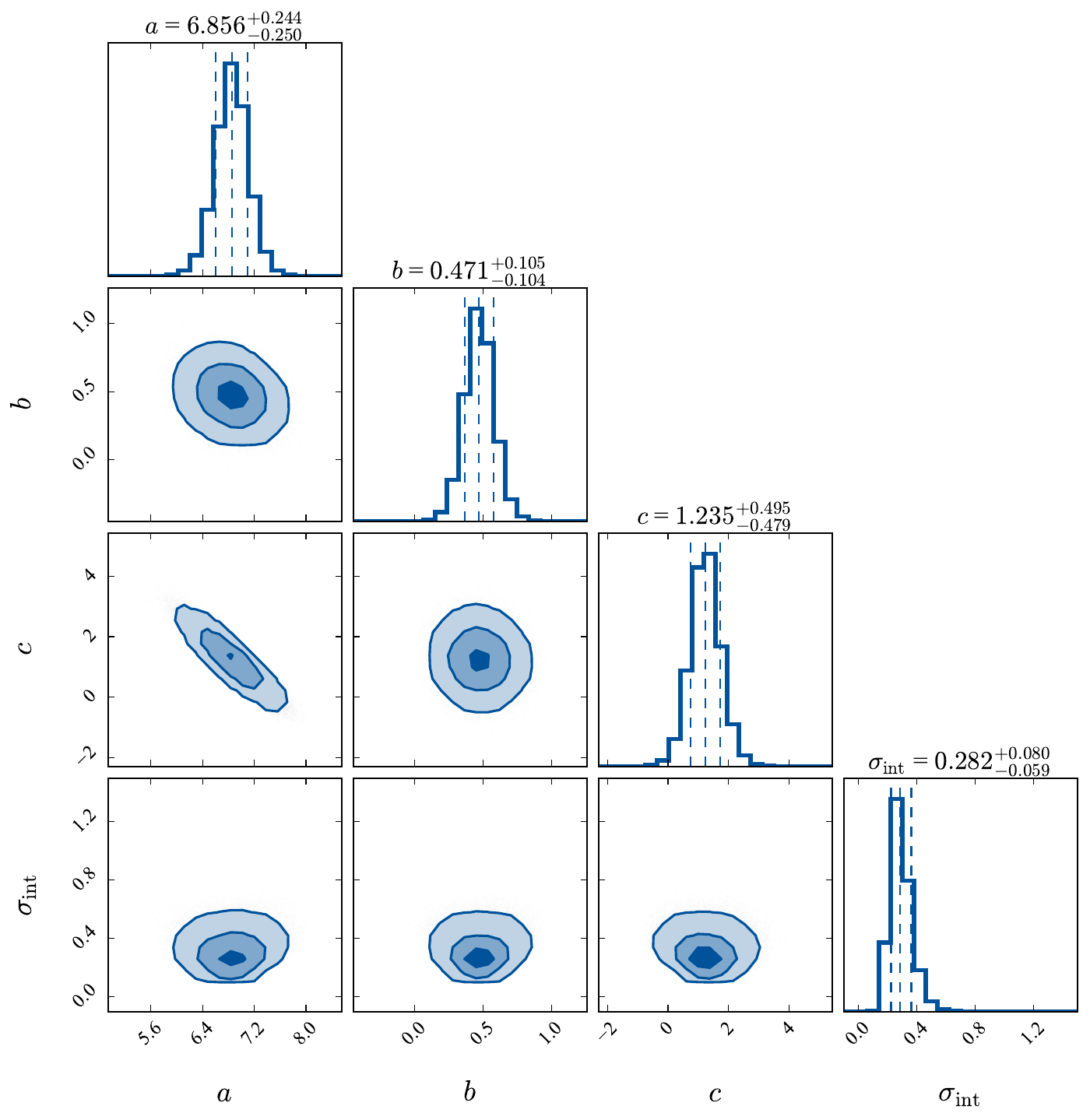}
\caption{Marginalized posterior distributions for the fiducial \pa\ estimator based on $L_{\rm P\alpha}$ and $\rm FWHM_{multi,P\alpha}$.
The fitted parameters are the normalization $a$, luminosity slope $b$, FWHM slope $c$, and intrinsic scatter $\sigma_{\rm int}$ in Eq.~\ref{eq:paschen_mbh_calibration}.
}
\label{mcmc_example}
\end{figure}

\begin{figure*}[t]
     \centering
     \subfigure[$\mathrm{FWHM_{multi}}\ \&\ L_{\rm P\alpha}${\label{sub_a_pa}}]{\includegraphics[width=0.42\textwidth]{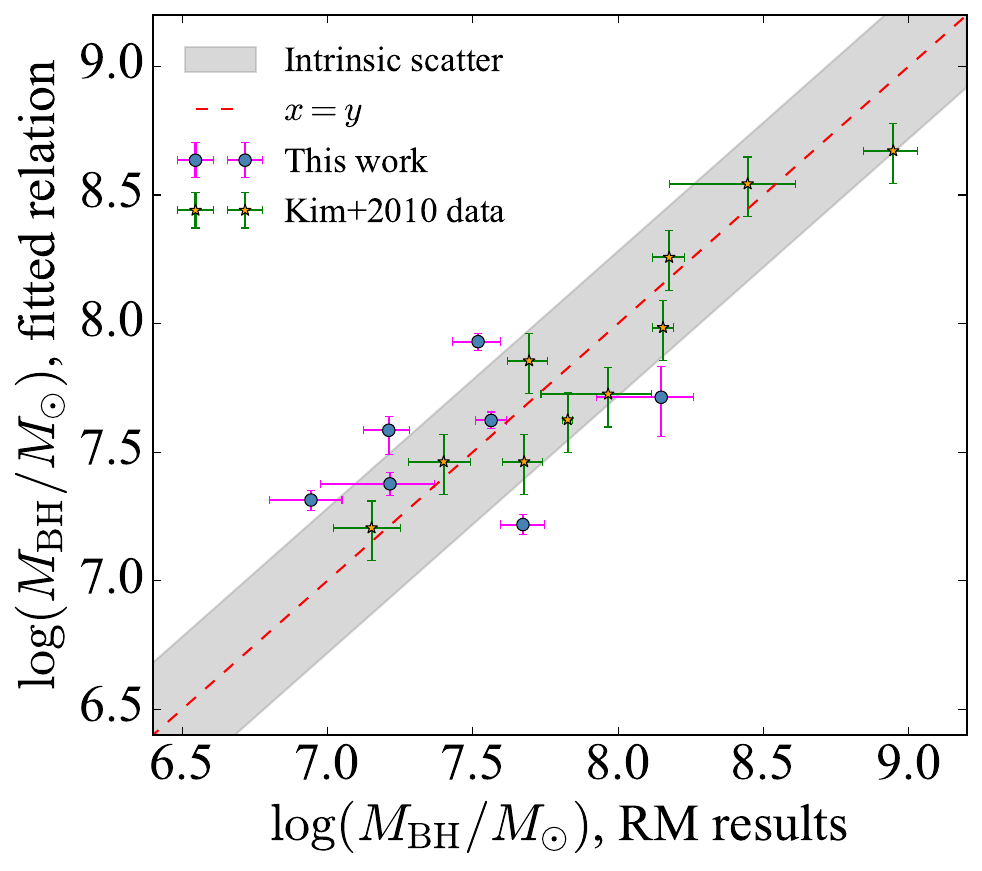}}\hspace{0.025\textwidth}
     \subfigure[$\mathrm{FWHM_{multi}}\ \&\ L_{5100}${\label{sub_b_pa}}]{\includegraphics[width=0.42\textwidth]{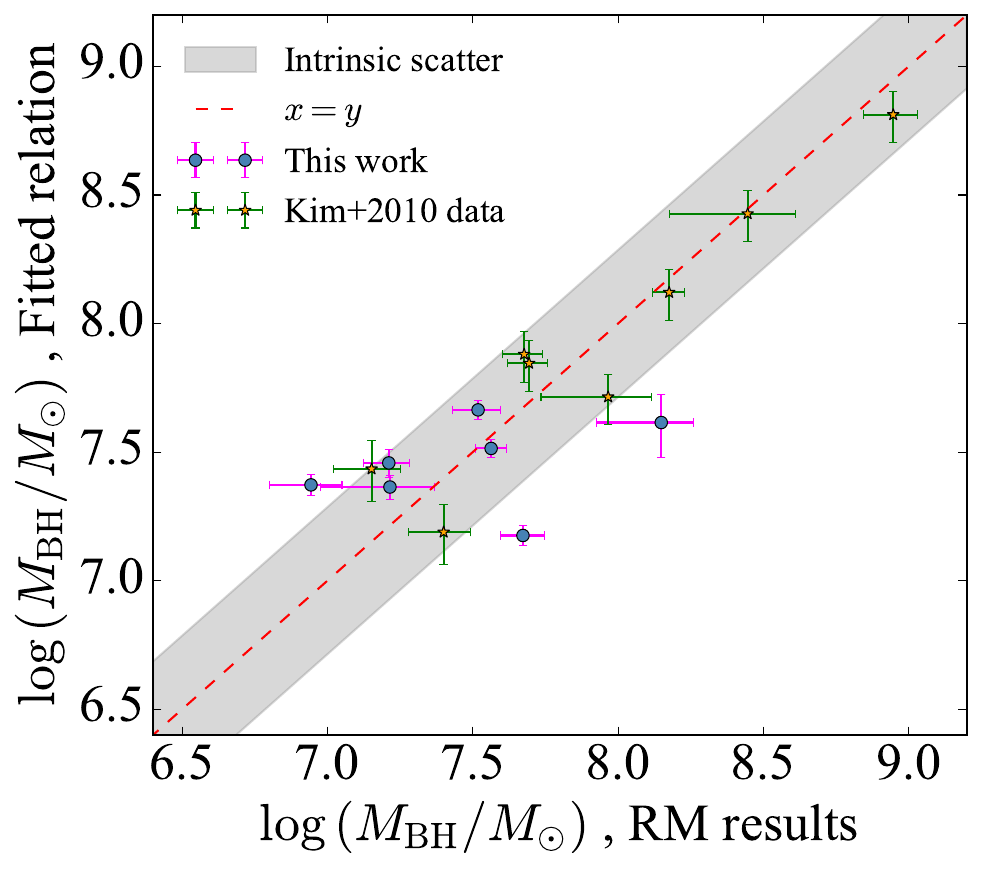}}\par\vspace{0.6ex}
     \subfigure[$\mathrm{FWHM_{\sigma}}\ \&\ L_{\rm P\alpha}${\label{sub_c_pa}}]{\includegraphics[width=0.42\textwidth]{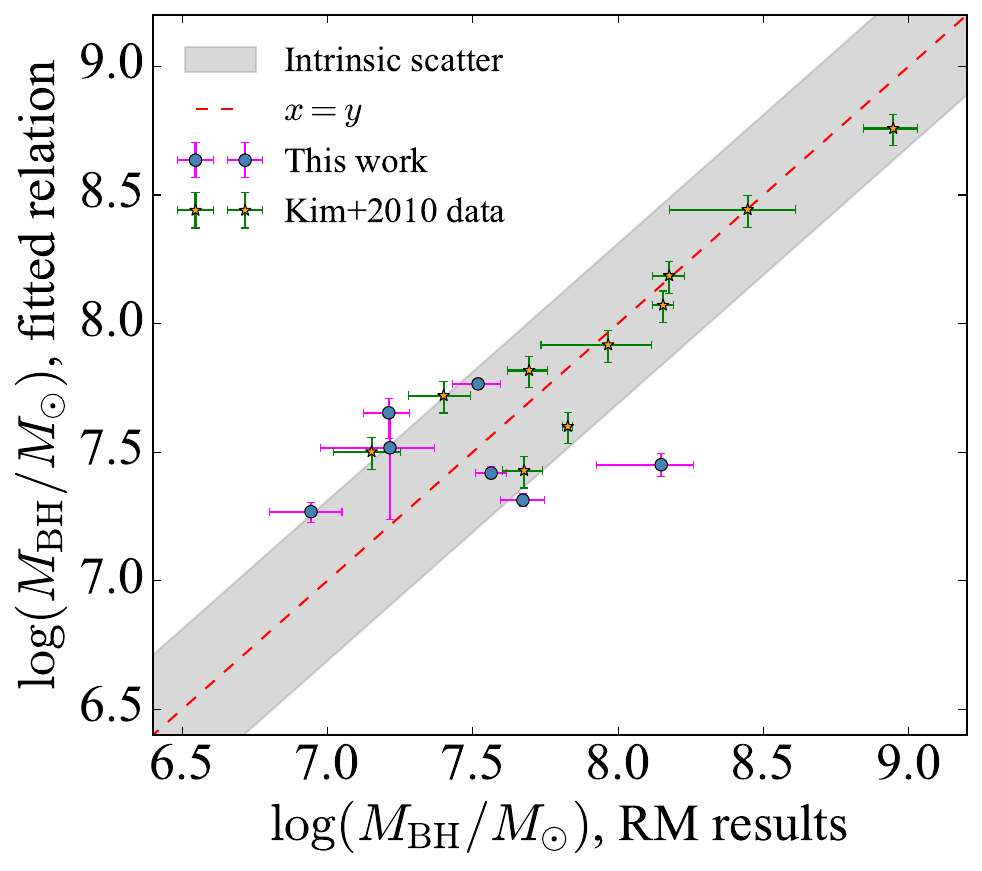}}\hspace{0.025\textwidth}
     \subfigure[$\mathrm{FWHM_{\sigma}}\ \&\ L_{5100}${\label{sub_d_pa}}]{\includegraphics[width=0.42\textwidth]{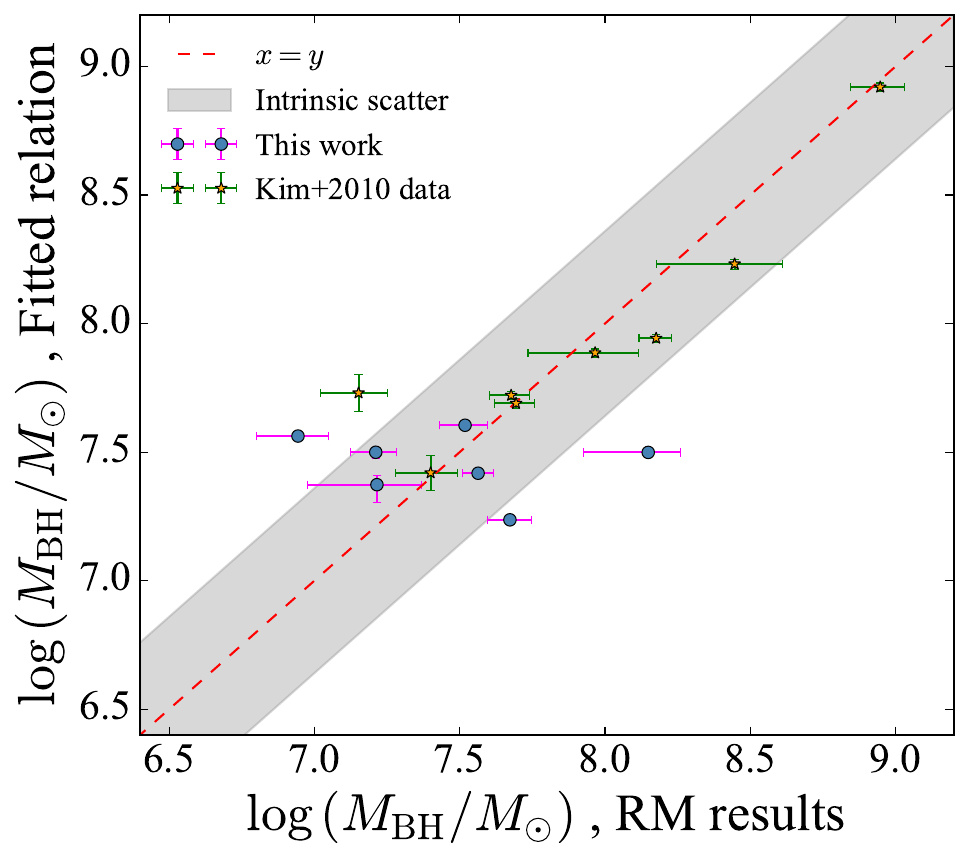}}
     \caption{\pa\ calibrations based on $\rm FWHM_{multi}$ (top) or $\rm FWHM_{\sigma}$ (bottom), combined with the Paschen line luminosity (left) or continuum luminosity $L_{5100}$ (right).}
    \label{Calib_pa}
\end{figure*}

\begin{figure*}[t]
     \centering
     \subfigure[$\mathrm{FWHM_{multi}}\ \&\ L_{\rm P\beta}${\label{sub_a_pb}}]{\includegraphics[width=0.42\textwidth]{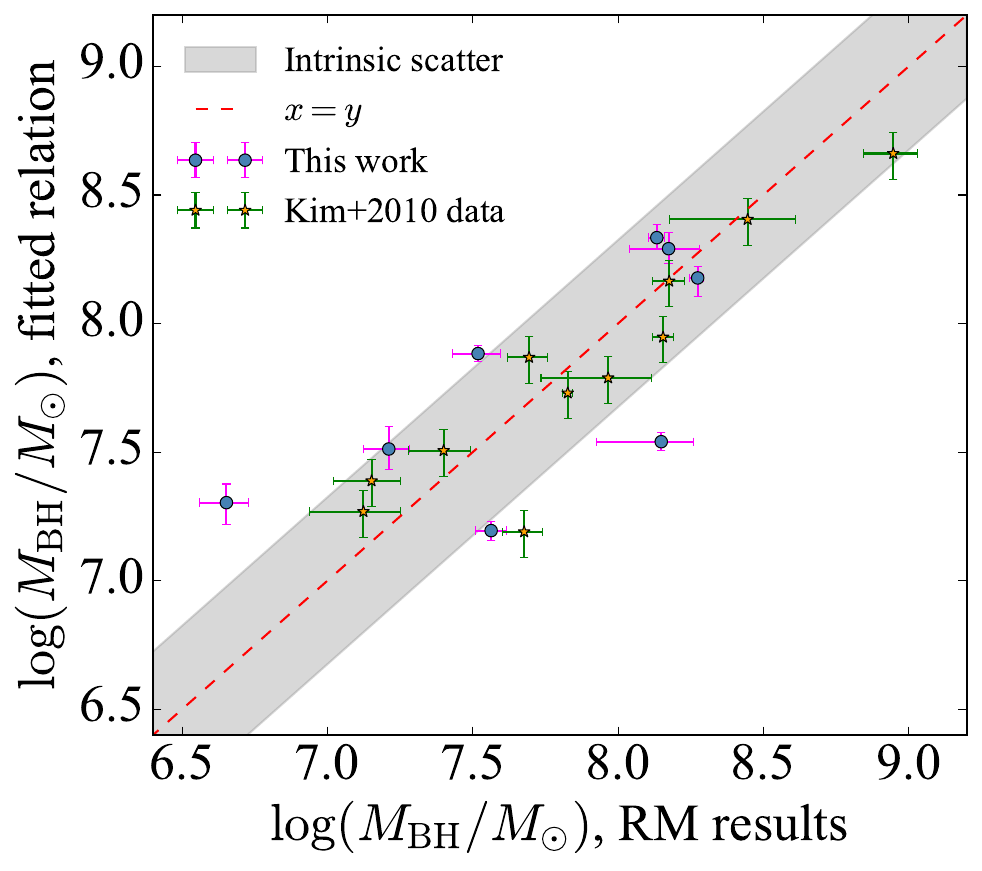}}\hspace{0.025\textwidth}
     \subfigure[$\mathrm{FWHM_{multi}}\ \&\ L_{5100}${\label{sub_b_pb}}]{\includegraphics[width=0.42\textwidth]{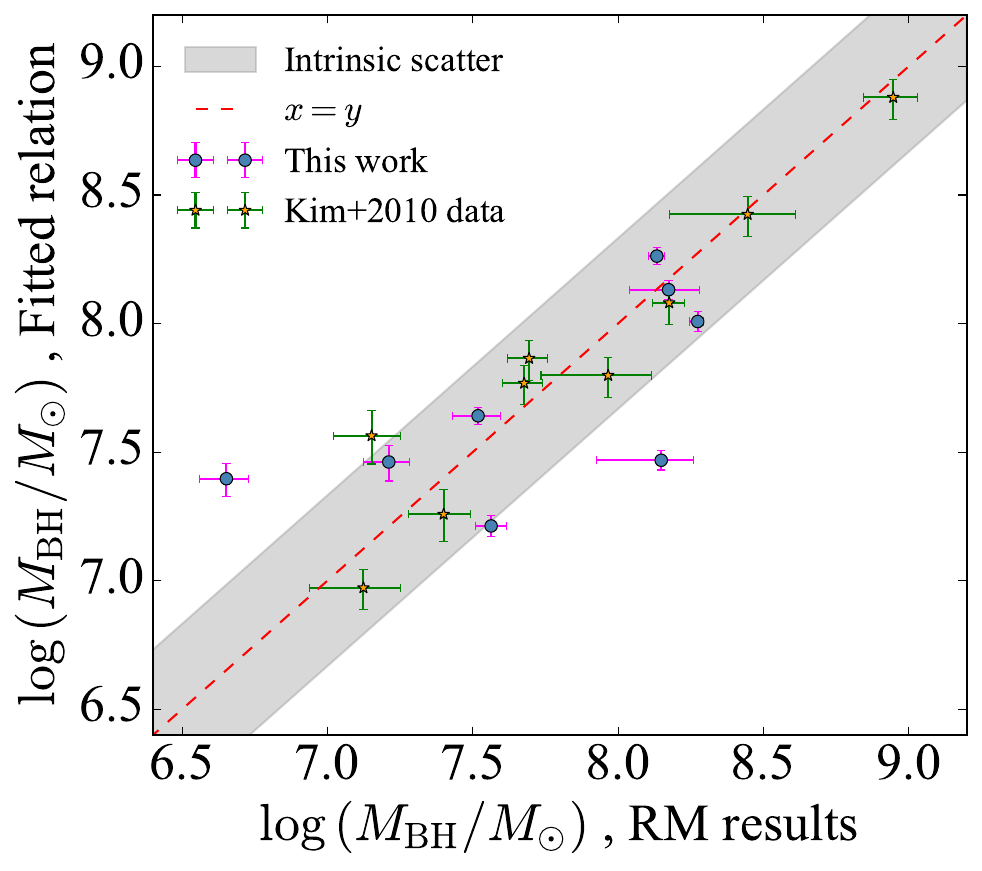}}\par\vspace{0.6ex}
     \subfigure[$\mathrm{FWHM_{\sigma}}\ \&\ L_{\rm P\beta}${\label{sub_c_pb}}]{\includegraphics[width=0.42\textwidth]{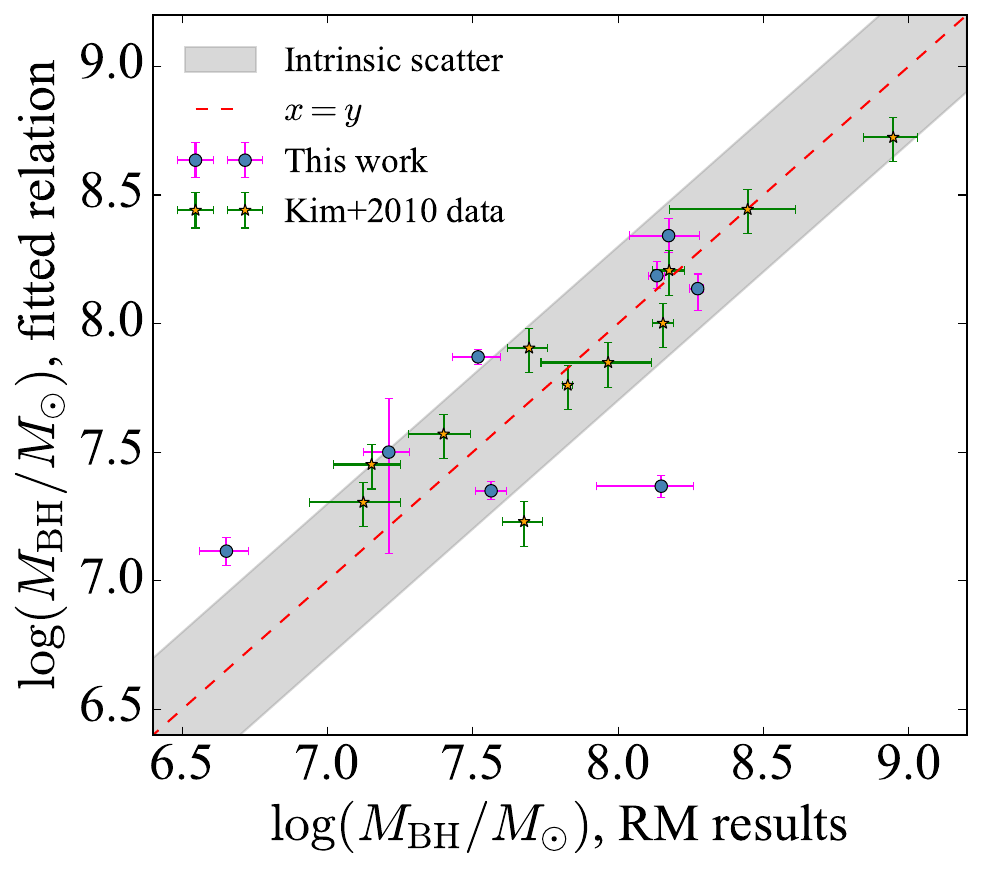}}\hspace{0.025\textwidth}
     \subfigure[$\mathrm{FWHM_{\sigma}}\ \&\ L_{5100}${\label{sub_d_pb}}]{\includegraphics[width=0.42\textwidth]{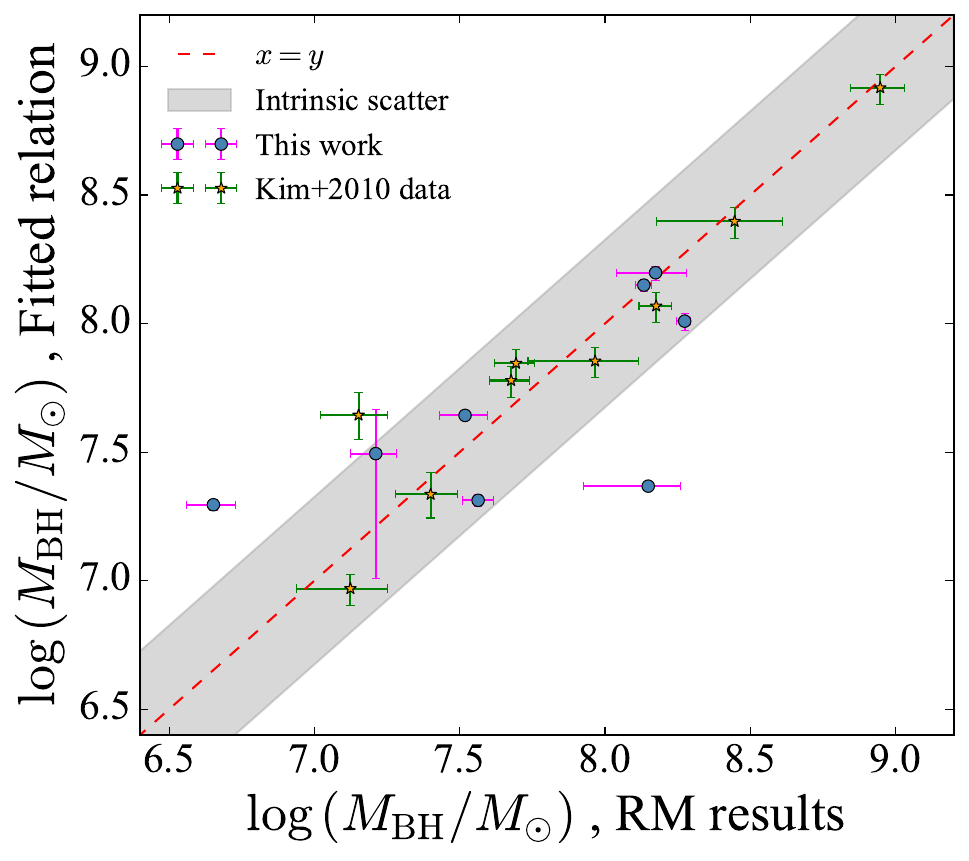}}
     \caption{Same as Fig.~\ref{Calib_pa}, but for \pb.}
    \label{Calib_pb}
\end{figure*}

The calibration sample combines the LBT measurements of SDSS-RM quasars with the \citet{Kim_2010} objects that have Paschen line measurements and RM masses.
Table~\ref{tab:paschen_params} lists measurements that satisfy the adopted fitting criteria; sources without an RM mass appear in the table but not in the calibration.
Table~\ref{tab:paschen_calibration} lists the resulting calibration parameters.
Figure~\ref{mcmc_example} shows an example posterior distribution, and Figures~\ref{Calib_pa} and~\ref{Calib_pb} compare the calibrated SE masses with the RM masses.

\begin{table*}[t]
\centering
\caption{Calibration parameters for SE black hole mass estimators based on the Paschen series.}
\label{tab:paschen_calibration}
\begingroup
\footnotesize
\setlength{\tabcolsep}{2.8pt}
\renewcommand{\arraystretch}{1.08}
\begin{adjustbox}{max width=0.94\textwidth}
\begin{tabular}{@{}lllcccc@{}}
\hline\hline
Line & FWHM type & Luminosity type & $a$ & $b$ & $c$ & Intrinsic scatter ($\sigma_{\rm int}$) \\
\hline
\multicolumn{7}{c}{Paschen-$\alpha$ estimators} \\
\hline
P$\alpha$ & Multi-component & Line & $6.86^{+0.24}_{-0.25}$ & $0.47^{+0.11}_{-0.10}$ & $1.24^{+0.49}_{-0.48}$ & $\textbf{0.28}^{+0.08}_{-0.06}$ \\
P$\alpha$ & Multi-component & Continuum & $5.46^{+0.43}_{-0.44}$ & $0.57^{+0.13}_{-0.13}$ & $1.12^{+0.55}_{-0.52}$ & $0.29^{+0.09}_{-0.07}$ \\
P$\alpha$ & Single Gaussian & Line & $7.27^{+0.15}_{-0.15}$ & $0.46^{+0.11}_{-0.11}$ & $0.52^{+0.33}_{-0.34}$ & $0.31^{+0.09}_{-0.06}$ \\
P$\alpha$ & Single Gaussian & Continuum & $5.80^{+0.48}_{-0.48}$ & $0.60^{+0.16}_{-0.16}$ & $0.17^{+0.42}_{-0.42}$ & $0.36^{+0.11}_{-0.08}$ \\
\hline
\multicolumn{7}{c}{Paschen-$\beta$ estimators} \\
\hline
P$\beta$ & Multi-component & Line & $7.00^{+0.24}_{-0.24}$ & $0.48^{+0.10}_{-0.09}$ & $0.91^{+0.46}_{-0.47}$ & $\textbf{0.32}^{+0.08}_{-0.06}$ \\
P$\beta$ & Multi-component & Continuum & $5.48^{+0.44}_{-0.44}$ & $0.60^{+0.12}_{-0.12}$ & $0.87^{+0.54}_{-0.53}$ & $0.33^{+0.09}_{-0.07}$ \\
P$\beta$ & Single Gaussian & Line & $7.07^{+0.17}_{-0.17}$ & $0.48^{+0.09}_{-0.09}$ & $0.86^{+0.32}_{-0.33}$ & $0.34^{+0.08}_{-0.06}$ \\
P$\beta$ & Single Gaussian & Continuum & $5.61^{+0.41}_{-0.40}$ & $0.60^{+0.12}_{-0.12}$ & $0.66^{+0.39}_{-0.39}$ & $0.33^{+0.09}_{-0.07}$ \\
\hline
\end{tabular}%
\end{adjustbox}
\tablefoot{
The fitted relation is given by Eq.~\ref{eq:paschen_mbh_calibration}.
The luminosity term is normalized to $10^{41}\ {\rm erg\ s^{-1}}$, and the FWHM term is normalized to $10^{3}\ {\rm km\ s^{-1}}$.
``Line'' refers to $L_{\rm P\alpha}$ or $L_{\rm P\beta}$, measured by integrating the fitted broad Paschen components after continuum subtraction.
The line luminosity rows use the photometrically corrected LBT flux scale, whereas the $L_{5100}$ rows are unchanged.
``Continuum'' refers to the rest-frame 5100~\AA\ AGN continuum luminosity.
``Multi-component'' refers to $\rm FWHM_{multi}$, measured from the combined broad Gaussian components, while ``Single Gaussian'' refers to $\rm FWHM_{\sigma}$, measured from a single-Gaussian fit to the Paschen profile.
}
\endgroup
\end{table*}

For \pa, our fiducial estimator uses the line luminosity $L_{\rm P\alpha}$ and the multi-component line width $\rm FWHM_{multi,P\alpha}$:
\begin{equation}
\begin{split}
\frac{M_{\rm BH}}{M_{\odot}} &=
10^{6.86^{+0.24}_{-0.25}}
\left( \frac{L_{\rm P\alpha}}{10^{41}\ {\rm erg\ s^{-1}}} \right)^{0.47^{+0.11}_{-0.10}}
\\
&\quad\times
\left( \frac{\rm FWHM_{multi,P\alpha}}{10^{3}\ {\rm km\ s^{-1}}} \right)^{1.24^{+0.49}_{-0.48}} ,
\end{split}
\end{equation}
with an intrinsic scatter of $0.28^{+0.08}_{-0.06}$ dex.

For \pb, the corresponding fiducial estimator is
\begin{equation}
\begin{split}
\frac{M_{\rm BH}}{M_{\odot}} &=
10^{7.00^{+0.24}_{-0.24}}
\left( \frac{L_{\rm P\beta}}{10^{41}\ {\rm erg\ s^{-1}}} \right)^{0.48^{+0.10}_{-0.09}}
\\
&\quad\times
\left( \frac{\rm FWHM_{multi,P\beta}}{10^{3}\ {\rm km\ s^{-1}}} \right)^{0.91^{+0.46}_{-0.47}} ,
\end{split}
\end{equation}
with an intrinsic scatter of $0.32^{+0.08}_{-0.06}$ dex.

The fitted intrinsic scatters range from 0.28 to 0.36 dex across the four combinations of luminosity and line width measurements, and their posterior intervals overlap. Among the models considered here, we adopt the combination of multi-component line width and line luminosity measurements as the fiducial calibration result because they represent the more comprehensive fitting and has smaller fitted intrinsic scatter. The single-Gaussian relations may be applicable when the spectral resolution or S/N is insufficient to constrain multiple broad component profile.
According to the simple photoionization scaling, $R_{\rm BLR}\propto L_{\rm ion}^{1/2}$, if the characteristic ionization parameter, gas density, and ionizing spectral shape are approximately independent of luminosity. And the measured H$\beta$ $R_{\rm BLR}$--$L_{5100}$ slope is also consistent with this scaling \citep{Bentz2013}.
Our fitted Paschen line luminosity slopes are consistent with $b \simeq 0.5$ within their uncertainties.
%although this comparison additionally assumes that the Paschen line luminosity traces the ionizing luminosity approximately linearly.
The posterior median FWHM slopes are below the virial expectation of 2, as also found in previous Paschen line calibrations in which all coefficients were allowed to vary freely \citep{Kim_2010, Kim_2015}; their interpretation should account for the current uncertainties and sample size.
Because our calibration is anchored to RM masses, the resulting coefficients need not be identical to those from calibrations tied partly to SE Balmer line masses.

\begin{figure*}[t]
\centering
\subfigure[\pa, fixed $b=0.5$ and $c=2${\label{sub_a_fixed}}]{\includegraphics[width=0.40\textwidth]{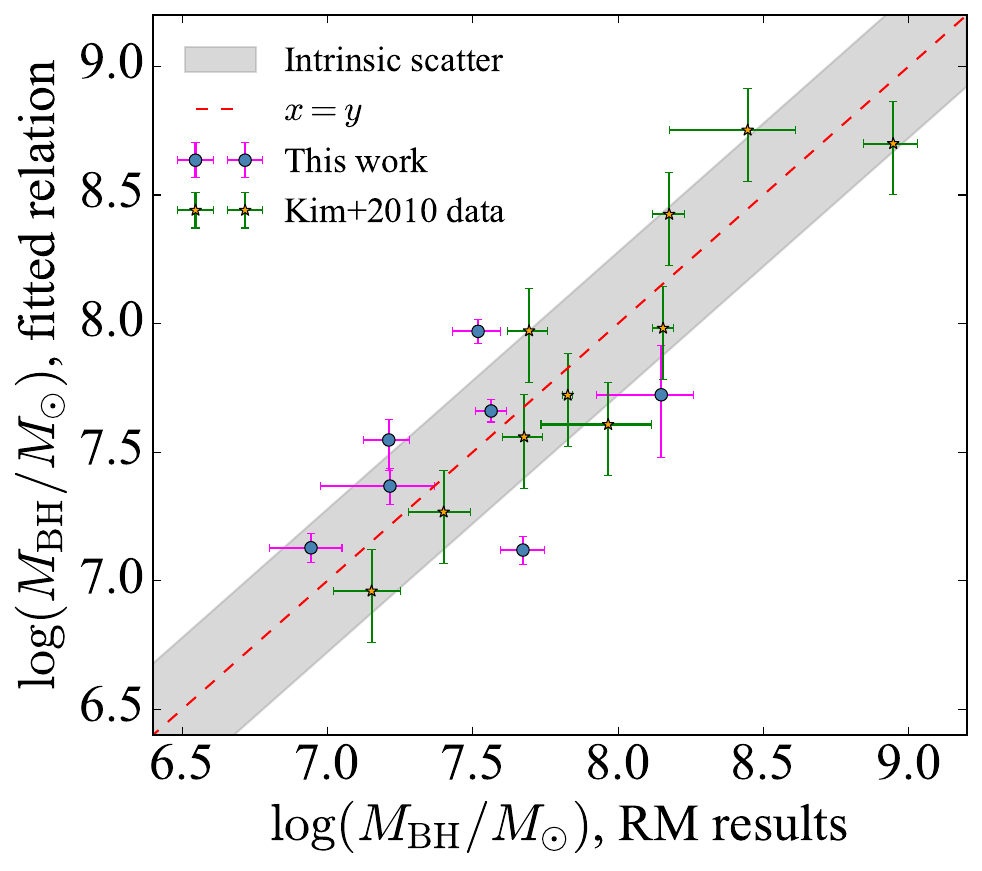}}\hspace{0.025\textwidth}
\subfigure[\pa, fixed $c=2${\label{sub_b_fixed}}]{\includegraphics[width=0.40\textwidth]{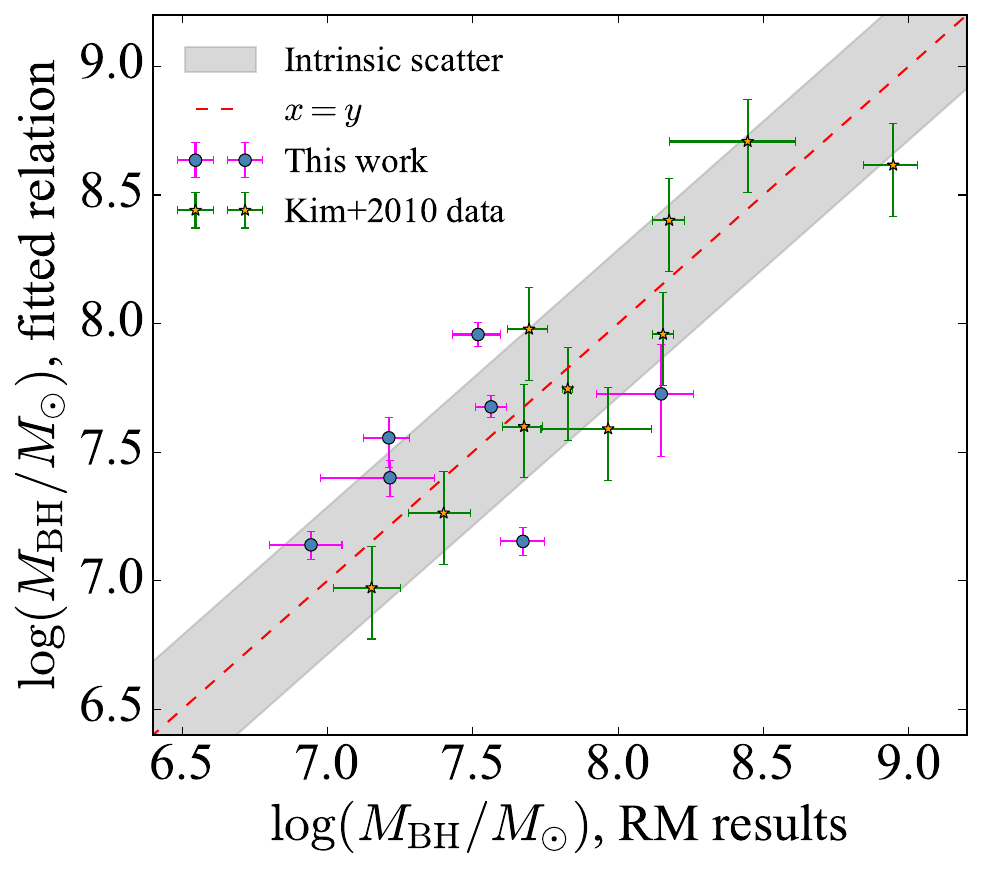}}\par\vspace{0.6ex}
\subfigure[\pb, fixed $b=0.5$ and $c=2${\label{sub_c_fixed}}]{\includegraphics[width=0.40\textwidth]{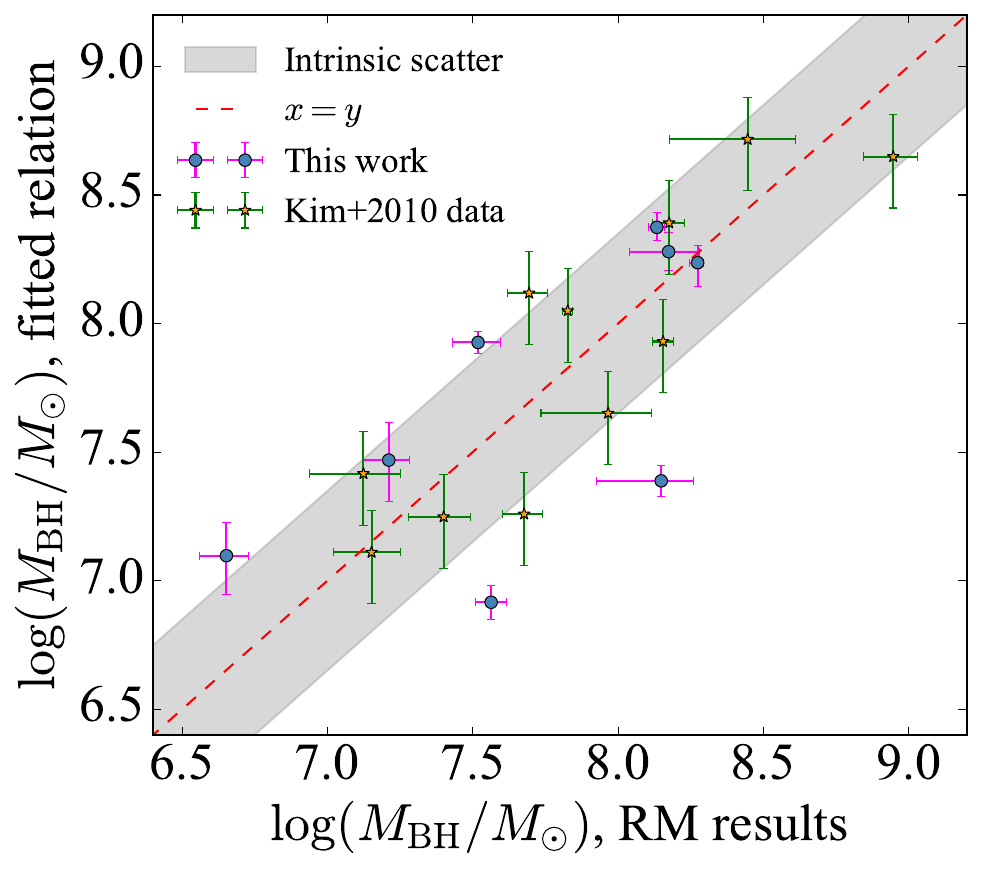}}\hspace{0.025\textwidth}
\subfigure[\pb, fixed $c=2${\label{sub_d_fixed}}]{\includegraphics[width=0.40\textwidth]{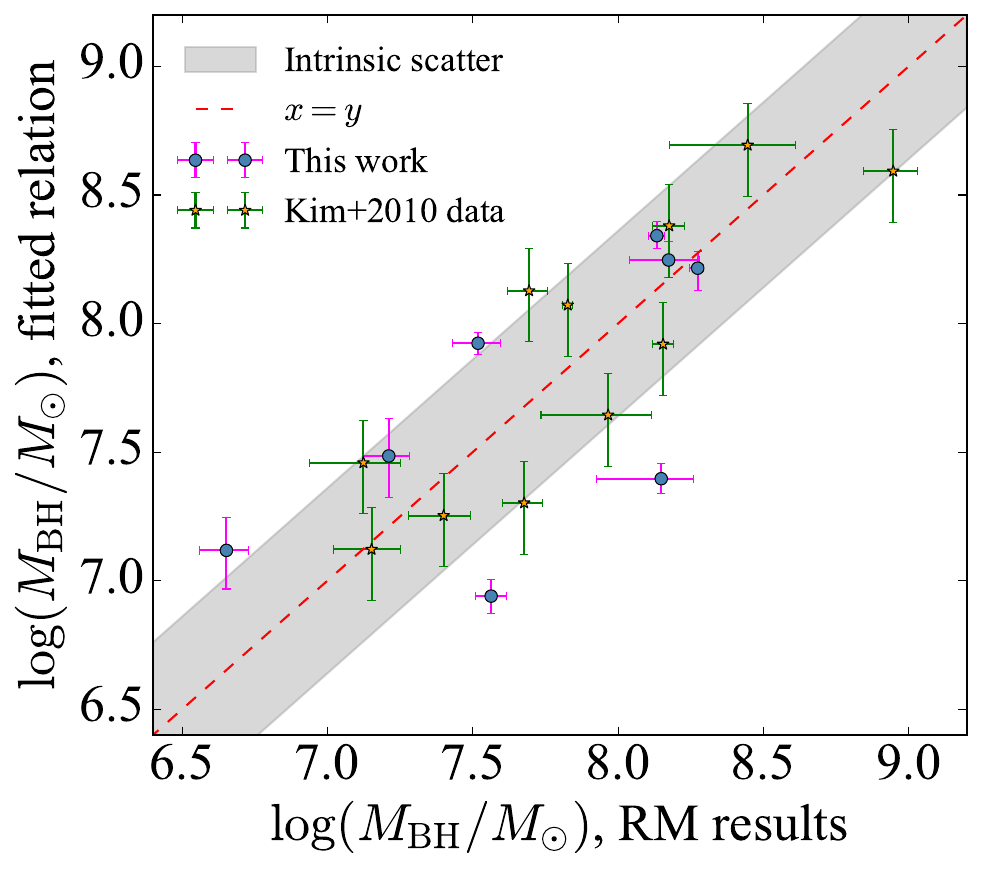}}
\caption{Constrained calibrations based on the Paschen line luminosity and $\rm FWHM_{multi}$. Panels \subref{sub_a_fixed} and \subref{sub_c_fixed} fix $b=0.5$ and $c=2$, while panels \subref{sub_b_fixed} and \subref{sub_d_fixed} fix only $c=2$. Blue circles show the LBT sample, orange stars show the archival \citet{Kim_2010} objects, gray bands show the fitted intrinsic scatter about the one-to-one relation, and red dashed lines mark equality.}
\label{fig:paschen_fixed_calibration}
\end{figure*}

We also fitted the multi-component line luminosity relations with either $b=0.5$ and $c=2$, or only $c=2$ fixed. We used the same calibration samples and likelihood and fitted $\sigma_{\rm int}$ in both cases.
Table~\ref{tab:paschen_fixed_calibration} lists the constrained fit parameters, and Figure~\ref{fig:paschen_fixed_calibration} shows the corresponding mass estimates. With $b=0.5$ and $c=2$, the fitted normalizations are $a=6.48^{+0.08}_{-0.08}$ for \pa\ and $a=6.48^{+0.09}_{-0.09}$ for \pb, with intrinsic scatters of $0.28^{+0.08}_{-0.06}$ and $0.35^{+0.08}_{-0.06}$ dex, respectively.
When only $c=2$ is fixed, the fitted luminosity slopes are $b=0.46^{+0.11}_{-0.11}$ for \pa\ and $b=0.47^{+0.11}_{-0.11}$ for \pb. Both are consistent with $b=0.5$ within their uncertainties. The corresponding intrinsic scatters are $0.29^{+0.08}_{-0.06}$ and $0.36^{+0.09}_{-0.07}$ dex.

\begin{table*}[t]
\centering
\caption{Alternative constrained multi-component Paschen line calibration parameters.}
\label{tab:paschen_fixed_calibration}
\begingroup
\small
\setlength{\tabcolsep}{7pt}
\renewcommand{\arraystretch}{1.08}
\begin{tabular}{llcccc}
\hline\hline
Line & Fixed slopes & $a$ & $b$ & $c$ & Intrinsic scatter ($\sigma_{\rm int}$) \\
\hline
P$\alpha$ & $b=0.5,\ c=2$ & $6.48^{+0.08}_{-0.08}$ & $0.50$ (fixed) & $2.00$ (fixed) & $0.28^{+0.08}_{-0.06}$ \\
P$\alpha$ & $c=2$ & $6.51^{+0.11}_{-0.11}$ & $0.46^{+0.11}_{-0.11}$ & $2.00$ (fixed) & $0.29^{+0.08}_{-0.06}$ \\
P$\beta$ & $b=0.5,\ c=2$ & $6.48^{+0.09}_{-0.09}$ & $0.50$ (fixed) & $2.00$ (fixed) & $0.35^{+0.08}_{-0.06}$ \\
P$\beta$ & $c=2$ & $6.50^{+0.13}_{-0.13}$ & $0.47^{+0.11}_{-0.11}$ & $2.00$ (fixed) & $0.36^{+0.09}_{-0.07}$ \\
\hline
\end{tabular}
\tablefoot{The fitted relation and normalizations follow Eq.~\ref{eq:paschen_mbh_calibration}. All fits use the Paschen line luminosity and $\rm FWHM_{multi}$. Fixed coefficients are not assigned posterior uncertainties; $\sigma_{\rm int}$ is fitted in every case. These relations are provided as alternative calibrated estimators with one or both virial slopes imposed.}
\endgroup
\end{table*}

The constrained and freely fitted relations differ in posterior median intrinsic scatter by at most $0.01$ dex for \pa\ and $0.04$ dex for \pb, and their posterior intervals overlap. The current data therefore do not establish a significant change in scatter when the slopes are fixed. We adopt the freely fitted relations as fiducial and provide the constrained relations as alternatives with conventional virial exponents. Both should be applied within the luminosity and line width ranges of the calibration sample.

\section{Discussion} \label{sec:discussion}

\subsection{Fiducial Paschen line mass estimators}

We adopt the Paschen line SE mass estimators based on line luminosity and multi-component FWHM as our fiducial relations.
For \pa, the fiducial relation uses the corrected $L_{\rm P\alpha}$ and $\rm FWHM_{multi,P\alpha}$ and has an intrinsic scatter of $0.28^{+0.08}_{-0.06}$ dex. For \pb, the corresponding relation has an intrinsic scatter of $0.32^{+0.08}_{-0.06}$ dex. 
These fitted scatters are within the range commonly associated with SE virial black hole mass estimates \citep[e.g.,][]{Vestergaard2006}. 
And they are smaller than those of typical black hole mass estimates, which are about 0.5 dex.
Direct comparisons should account for differences in sample size, dynamic range, and calibration method.
These results indicate that Paschen lines can serve as SE mass tracers when their broad profiles are well measured.

We also provide alternative calibrations in which $c=2$, or both $b=0.5$ and $c=2$, are fixed. Their posterior median intrinsic scatters differ from those of the freely fitted fiducial relations by no more than 0.04 dex, and the posterior intervals overlap. The freely fitted relations provide an empirical description of the present sample, whereas the constrained relations may be useful when a conventional virial scaling or a common set of slopes is required.

Our calibrated relations only use RM-based black hole masses as the reference. However, these relations still remain subject to scatter in the $R-L$ relation, line width measurement uncertainties, the adopted virial factor, and BLR structure and kinematics \citep[e.g.,][]{Peterson2004,Vestergaard2006}. Anchoring the calibration to RM masses avoids an intermediate calibration against Balmer line SE masses, but does not remove uncertainties in the RM mass scale and virial factor.

Previous studies developed Paschen line virial mass relations using NIR spectroscopy \citep[e.g.,][]{Kim_2010,Kim_2015,LaFranca2015,Ricci2017}. Our analysis extends this work with LBT/LUCI spectra of SDSS-RM quasars and calibrates the \pa\ and \pb\ relations against RM-based masses. These relations can be applied to reddened and obscured quasars when suitable Paschen measurements are available.

For reddened or obscured quasars, the line luminosity versions may be more directly applicable than the continuum-based versions because both input quantities are measured from the same NIR spectrum.
The optical continuum can be affected by dust attenuation, host galaxy contamination, and differences between the epochs of the optical and NIR observations.
The continuum-based relations have similar fitted scatter and may be used when an $L_{5100}$ measurement is available, after correction.

The multi-component and single-Gaussian FWHM relations have consistent fitted scatters within the posterior uncertainties. We use the multi-component measurements as fiducial because they describe the overall line profile better.
Paschen emission line profiles can contain both a narrow core and an extended broad base, so a single Gaussian cannot always capture the full line structure. 
The single-Gaussian estimators yield intrinsic scatters of $\simeq0.3$ dex.
They may be applicable to lower-S/N or lower-resolution spectra, where a detailed decomposition of the broad line profile is not well constrained.

\subsection{Comparison with literature results}

We compare our fiducial estimators with Eqs.~11 and 12 of \citet{Kim_2010}, which are based on \pa\ and \pb\ line luminosities and FWHMs. For each object, we compute both SE masses from the same Paschen line luminosity and $\rm FWHM_{multi}$. The comparison includes the LBT sample and the archival objects that also enter our calibration.

\begin{figure*}[t]
\centering
\includegraphics[width=0.74\textwidth]{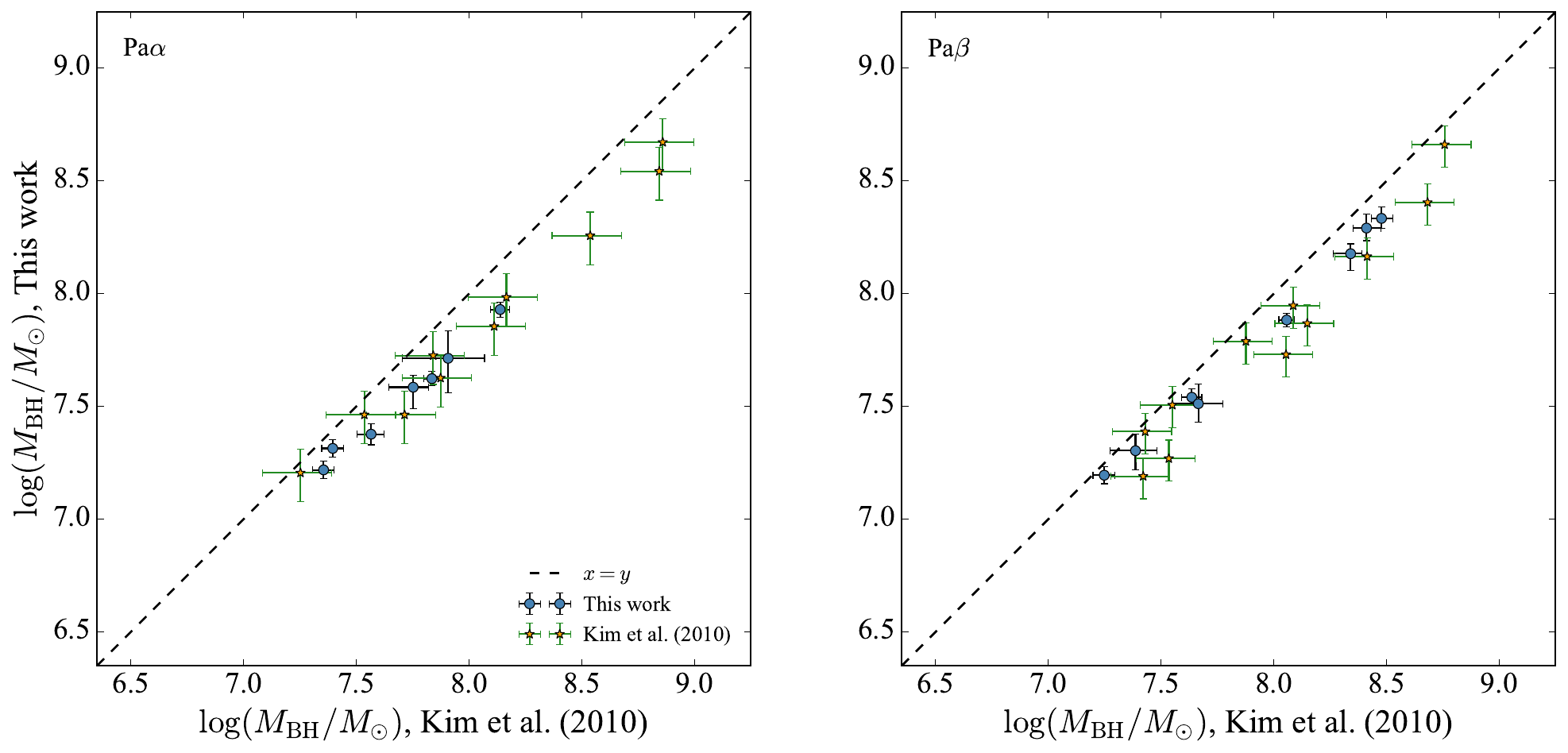}
\caption{
Comparison between SE black hole masses from this work and \citet{Kim_2010}, computed from the same Paschen line luminosity and $\rm FWHM_{multi}$.
The left and right panels show the \pa\ and \pb\ estimators. Blue circles denote the LBT sample, orange stars denote the archival calibration objects, and the dashed line marks the one-to-one relation.
}
\label{fig:paschen_kim_comp}
\end{figure*}

Figure~\ref{fig:paschen_kim_comp} compares the two sets of mass estimates over $\log(M_{\rm BH}/M_\odot) \sim 7$--$9$.

For the LBT sample, our relations give masses that are lower than the \citet{Kim_2010} predictions by $0.17$ dex for \pa\ and $0.12$ dex for \pb\ on average.
Both comparisons show some variation in the offset with mass over the sampled range, with a stronger trend for \pa.
The offsets can reflect differences in the fitted coefficients, calibration samples, and mass anchors.
In particular, our relations are calibrated against RM-based black hole masses, whereas the \citet{Kim_2010} calibration also used black hole masses inferred from Balmer line SE estimators.

The mean offsets are smaller than the intrinsic scatters of the corresponding fiducial relations. Because the archival objects from \citet{Kim_2010} contribute to our fit, this comparison tests consistency between calibration prescriptions rather than providing an independent validation.

\citet{Jiang2026COSMOS3D} used JWST COSMOS-3D slitless spectroscopy to identify AGNs with broad Paschen lines at cosmic noon, calibrate Paschen-based SE mass estimators, and derive Paschen line luminosity functions. Their sample contains 44 \pa\ AGNs at $z=1.1$--$1.7$ and 18 \pb\ AGNs at $z=2.1$--$2.8$. Nineteen \pa\ and five \pb\ sources with Mg\,{\sc ii} spectroscopy from the Dark Energy Spectroscopic Instrument form their calibration sample.

Their sample spans $\log L_{\rm Paschen}\simeq41.3$--$43.3$ and $\log(M_{\rm BH}/M_\odot)\simeq7.0$--$9.5$. After the photometric correction, our LBT sample spans $\log L_{\rm P\alpha}\simeq40.81$--$41.96$, $\log L_{\rm P\beta}\simeq41.00$--$42.88$, and $\log(M_{\rm BH}/M_\odot)\simeq6.65$--$8.27$ at $z\simeq0.12$--$0.72$; including the archival objects extends our calibration to $\log L_{\rm Paschen}\simeq43.7$ and $\log(M_{\rm BH}/M_\odot)\simeq8.95$. The two studies therefore overlap but emphasize different regions of luminosity, black hole mass, and redshift.

When all coefficients are fitted, \citet{Jiang2026COSMOS3D} obtain \pa\ luminosity and FWHM exponents of $b=0.44\pm0.10$ and $c=1.96\pm0.27$, respectively. Their luminosity exponent is consistent with our $b=0.47^{+0.11}_{-0.10}$. Their posterior median FWHM exponent is higher than our $c=1.24^{+0.49}_{-0.48}$, although the uncertainty intervals overlap. An equivalent \pb\ comparison with all coefficients fitted is unavailable because they derive the \pb\ estimator by scaling their \pa\ relation, whereas we fit the two lines independently. The difference in the \pa\ coefficients may reflect the calibration strategies and parameter ranges: they use Mg\,{\sc ii} SE masses as the reference and a Paschen-selected sample with JWST-based host and dust modeling, while we use RM-based masses for primarily optically selected quasars. Both studies measure the FWHM from the combined multi-Gaussian broad line profile.

When the FWHM exponent is fixed to the virial value of $c=2$, the fitted luminosity slopes agree within their uncertainties. Using our luminosity normalization, the \citet{Jiang2026COSMOS3D} relations have $(a,b,c)=(6.79,0.44,2)$ for \pa\ and approximately $(6.54,0.44,2)$ for \pb, compared with $(6.51,0.46,2)$ and $(6.50,0.47,2)$ in this work.

\subsection{Sources without prominent broad Paschen emission} \label{sec:paschen_nondetections}

A small number of SDSS-RM quasars show broad Balmer emission in the optical spectra but no broad \pa\ or \pb\ component meeting our detection criteria in the LBT spectra.
Dust attenuation alone would not generally be expected to suppress Paschen emission more strongly than Balmer emission. NIR studies have also reported the complementary case in which broad Paschen emission is detected in sources without broad optical Balmer lines \citep[e.g.,][]{Lamperti2017,denBrok2022}.

One possible explanation is that the broad Paschen component has low contrast relative to the continuum in some spectra.
Paschen lines can be weaker than the Balmer lines, and their broad flux can be distributed over several thousand ${\rm km~s^{-1}}$. The narrow Paschen core may therefore remain detectable when the broad base is below the effective detection threshold.
NIR sky line residuals, telluric absorption, and uncertainties in continuum placement can further affect broad, low-contrast components.

Long-term variability may also contribute. 
The optical SDSS-RM spectra and the LBT NIR spectra are not simultaneous, and broad line strengths can change in response to continuum variability.
In extreme cases, changing-look quasars show the disappearance or strong weakening of broad emission lines over multi-year timescales \citep[e.g.,][]{MacLeod2016,Ruan2016}. 
A different continuum state during the LBT observations could therefore contribute to differences between the optical and Paschen profiles.

The present data do not distinguish among weak broad Paschen emission, limited sensitivity in the relevant NIR spectral regions, uncertainty in continuum placement, and non-simultaneous variability.
Contemporaneous optical and NIR spectroscopy with higher S/N would be needed to determine whether these sources have unusual ratios of Paschen to Balmer line emission or broad Paschen components below the detection threshold of the current LBT data.

\flushbottom
\section{Summary} \label{sec:summary}

We present RM-based SE black hole mass estimators using broad Paschen \pa\ and \pb\ emission. The LBT/LUCI spectra provide photometrically corrected Paschen line luminosities and both multi-component and single-Gaussian FWHM measurements; archival SDSS-RM spectra provide $L_{5100}$ for the continuum-based alternatives.

Our fiducial relations based on line luminosity and multi-component FWHM have intrinsic scatters of $0.28^{+0.08}_{-0.06}$ dex for \pa\ and $0.32^{+0.08}_{-0.06}$ dex for \pb. Relations with $c=2$, or with $b=0.5$ and $c=2$, differ in posterior median scatter by at most 0.04 dex. The single-Gaussian relations and those based on continuum luminosity provide alternatives when the corresponding measurements are more appropriate for the available data.

For the LBT sample, the mean offsets from the \citet{Kim_2010} predictions are $-0.17$ dex for \pa\ and $-0.12$ dex for \pb. Because their archival objects also enter our calibration, this comparison is a consistency check rather than an independent validation.

Several quasars show broad Balmer emission but no broad Paschen component meeting our criteria. The present data do not distinguish among weak Paschen emission, NIR residuals, uncertainty in continuum placement, and non-simultaneous variability; contemporaneous higher-S/N spectroscopy would help test these possibilities. Within the parameter range and limitations of the present sample, the calibrated relations provide an additional option for estimating masses when optical or ultraviolet broad line measurements are affected by dust attenuation or unavailable.

\begin{acknowledgements}
We thank Eduardo Ba\~nados for helpful discussions during the development of this work.
SS and LJ acknowledge support from the National Science Foundation of China (12225301).
SEIB, KP, and BS are supported by the Deutsche Forschungsgemeinschaft (DFG) under Emmy Noether grant number BO 5771/1-1a.

The LBT is an international collaboration among institutions in the United States and Europe. At the time data were acquired for this research, LBT Corporation Members were the University of Arizona on behalf of the Arizona Board of Regents; Istituto Nazionale di Astrofisica (INAF), Italy; LBT Beteiligungsgesellschaft, Germany, representing the Max-Planck Society, the Leibniz Institute for Astrophysics Potsdam, and Heidelberg University; and The Ohio State University, representing The Ohio State University, University of Notre Dame, University of Minnesota, and University of Virginia. This research used the facilities of the Italian Center for Astronomical Archive (IA2) operated by INAF at the Astronomical Observatory of Trieste. Observations have benefited from the use of ALTA Center (alta.arcetri.inaf.it) forecasts performed with the Astro-Meso-Nh model. Initialization data of the ALTA automatic forecast system come from the General Circulation Model (HRES) of the European Centre for Medium Range Weather Forecasts.

The observations used the LBT/LUCI and SDSS facilities. This research made use of Astropy \citep{2013A&A...558A..33A,2018AJ....156..123A,2022ApJ...935..167A}, PypeIt, PyQSOFit, emcee, and matplotlib \citep{Hunter:2007}.
This publication makes use of data products from the Spectro-Photometer for the History of the Universe, Epoch of Reionization and Ices Explorer (SPHEREx), which is a joint project of the Jet Propulsion Laboratory and the California Institute of Technology, and is funded by the National Aeronautics and Space Administration.
\end{acknowledgements}

\bibliographystyle{bibtex/aa}
\bibliography{sample701}

\end{document}